\documentclass[%
 aip,
 amsmath,amssymb,
 reprint,%
]{revtex4-1}

\usepackage{graphicx}% Include figure files
\usepackage{dcolumn}% Align table columns on decimal point
\usepackage{bm}% bold math
\usepackage{float}
\usepackage{soul, xcolor}
\usepackage[utf8]{inputenc}
\usepackage[T1]{fontenc}
\usepackage{mathptmx}
\usepackage{etoolbox}

\makeatletter
\def\@email#1#2{%
 \endgroup
 \patchcmd{\titleblock@produce}
  {\frontmatter@RRAPformat}
  {\frontmatter@RRAPformat{\produce@RRAP{*#1\href{mailto:#2}{#2}}}\frontmatter@RRAPformat}
  {}{}
}%
\makeatother
\begin{document}

\preprint{AIP/123-QED}

\title[Cavitation Onset in an Impulsively Accelerated Liquid Column]{Cavitation Onset in an Impulsively Accelerated Liquid Column\\}
% Force line breaks with \\
\author{T. Sobral}
 \affiliation{Department of Mechanical Engineering, McGill University, Montréal, Québec H3A 0C3, Canada.}%Lines break automatically or can be forced with \\
 \email{taj.sobral@mail.mcgill.ca}
 \author{J. Kokkalis}
 \affiliation{Department of Mechanical Engineering, McGill University, Montréal, Québec H3A 0C3, Canada.}%Lines break automatically or can be forced with \\
\author{K. Romann}
 \affiliation{Department of Mechanical Engineering, McGill University, Montréal, Québec H3A 0C3, Canada.}%Lines break automatically or can be forced with \\
\author{J. Nedić}
 \affiliation{Department of Mechanical Engineering, McGill University, Montréal, Québec H3A 0C3, Canada.}%Lines break automatically or can be 
 \author{A. J. Higgins}
 \affiliation{Department of Mechanical Engineering, McGill University, Montréal, Québec H3A 0C3, Canada.}

\date{\today}% It is always \today, today,

\begin{abstract}
This paper introduces a novel piston-driven apparatus to study the onset of cavitation in an impulsively accelerated liquid column as it compresses a closed gas volume. The experiment is monitored using high-speed videography and piezoelectric pressure transducers. Cavitation onset is observed in the liquid column as it undergoes an abrupt deceleration and is associated with a sudden drop in pressure in the liquid that leads to negative pressure (tension). A novel numerical modeling approach is introduced where the liquid column is treated as a spring-mass system. This approach can reproduce compressibility effects in the liquid column and is used to investigate the wave dynamics responsible for the onset of tension and cavitation in the liquid column. The model is formulated as a coupled set of non-linear differential equations that reproduce the dynamics of an experiment while capturing the pressure wave activity in the liquid column. A parametric study is conducted experimentally and numerically to investigate the behavior behind the onset of cavitation. The mechanism for the onset of cavitation is identified as a series of wave reflections at the boundaries of the liquid column, and this mechanism is found to be well reproduced by the model. While a traditional cavitation number criterion is shown to be unable to predict cavitation onset in our experiment, our numerical model is found to correctly predict the onset of cavitation for a wide range of experimental parameters.
\end{abstract}

\maketitle

\section{\label{sec:Introduction}Introduction}
Liquids can vaporize through two distinct physical processes: increasing the temperature above the saturation vapor/liquid temperature or decreasing the pressure below the vapor pressure\cite{Brennen_2013}. While the former is the commonly known phenomenon of \textit{boiling}, the latter is common only in particular engineering applications and some natural phenomena and is known as \textit{cavitation}. From a thermodynamic point of view, these two processes are almost equivalent. However, most research has been focused specifically on understanding cavitation due to its ubiquity in many hydrodynamic phenomena. Applications include hydraulic propellers and pumps\cite{Brennen_2011}, underwater explosions\cite{Cole_1965}, medical imagery and treatment\cite{Roovers_2019}, and auto-injector devices\cite{Veilleux_2018}.

Early research on cavitation dates back to the late 1800s, with cavitation in the wake of ship propellers causing a drop in performance and significant damage to the propeller\cite{Trevena_1984, Brennen_2013}. It is now well known that cavitation may occur in any liquid flow where the speed is such that the local static pressure drops below the vapor pressure\cite{Brennen_2011, Brennen_2013, Knapp_1958}. In practice, this can occur in the wake of propellers, behind any fast-moving underwater object, and in Venturi tubes\cite{Knapp_1958}. In such flows, it is common to introduce the following cavitation number\cite{Brennen_2011, Brennen_2013, Pan_2017} based on the dynamic pressure drop in a flowing liquid:
\begin{equation}
    \mathrm{\sigma} = \frac{p_1-p_\mathrm{v}}{\frac{1}{2}\,\rho_\ell\, U^2},
    \label{eq: Ca1}
\end{equation}
where $\rho_\ell$ is the density of the liquid, $p_\mathrm{v}$ is its vapor pressure, $U$ is the velocity of the flow, and $p_1$ is the reference static pressure. Although the threshold value of $\sigma$ below which the flow will cavitate varies across applications, it is generally understood\cite{Pan_2017, Fatjo_2016} that cavitation will occur where $\sigma \ll 1$. 

Beyond these engineering applications, much research has been conducted on the physics of liquids in tension\cite{Berthelot_1849, Trevena_1978, Frenkel_1955, Bull_1956, Knapp_1958, Skripov, Couzens_1974, Trevena_1984, Richards_1980, JOSEPH_1998}. Theoretical studies\cite{JOSEPH_1998, Skripov, Frenkel_1955} of pure liquids found that liquids should be able to sustain tensions on the order of $10^4$~bar. This is well beyond what has been seen experimentally, with the first experimental study conducted in 1849 by Berthelot yielding values of up to 50~bar\cite{Berthelot_1849, Trevena_1978}. 

Since then, many other experimental studies\cite{Trevena_1984, Couzens_1974, Richards_1980, Bull_1956, Williams_2004} have shown that liquids can undergo values of tension on the order of $1$--$10^3$~bar. This is commonly done experimentally by generating a shock directed toward a free surface that reflects as a rarefaction wave, putting the liquid in a state of tension. This has been done through the use of a detonation tube\cite{Richards_1980}, a \textit{bullet-piston} apparatus\cite{Bull_1956, Trevena_1984, Couzens_1974, Williams_2004}, and more recently on silicone oils using plate-impact experiments\cite{Huneault_2019}. Although many studies have been conducted to quantify the tension that liquids can sustain before cavitating, significant discrepancies still exist in the literature\cite{Huneault_2019, Williams_2004}.
This disagreement is generally explained by arguments on the purity of the fluid\cite{Brennen_2013}, the adhesive force between the liquid and the wall\cite{Trevena_1978, Trevena_1984}, or the rate of stress that the liquid is subjected to\cite{Trevena_1984}. In particular, solid or gaseous inclusions in the liquid can act as nucleation sites that decrease the threshold tension for the liquid to cavitate\cite{Brennen_2013, Knapp_1958, Trevena_1984}.
\begin{figure*}[]
        \includegraphics[width = \textwidth]{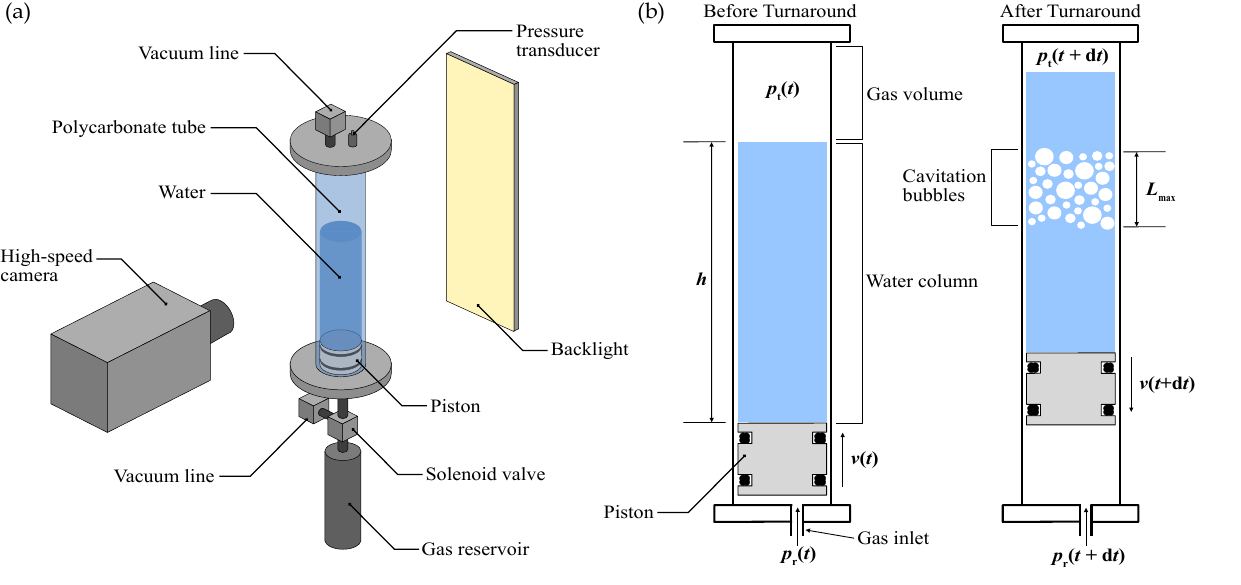}
\caption{\label{fig: Exp_Ap} Experimental apparatus. (a) Schematic of experimental apparatus, (b) schematic representation of the experiment test section showing key behavior and parameters (dimensions not to scale).}
\end{figure*}

Recently, some experimental studies have been interested in the onset of cavitation caused by acceleration in liquid columns\cite{Fatjo_2016, Pan_2017, Xu_2020, Chen_2004, Veilleux_2018}. These experiments, though qualitatively similar to the shock tube and bullet-piston methods, differ from them in that the accelerations involved are too small to generate shock waves. One common method, the \textit{tube-arrest} method, consists in partially filling a container with liquid and accelerating the container downward by a sharp impact, thus creating tension directly by forcing the bottom wall away from the liquid. Studies using this technique found that cavitation occurs at the bottom interface between the liquid and the solid wall of the container\cite{Pan_2017, Chen_2004}. 

Other methods such as drop-tube and surface piston impact are more analogous to the bullet-piston experiment. They consist in sending a compressive pulse through the liquid that eventually reflects off of a free surface as a rarefaction wave that puts the liquid in a state of tension\cite{Veilleux_2018, Pan_2017}. Results with these two methods generally show cavitation in the bulk of the liquid, not limited to the liquid-solid interface\cite{Veilleux_2018, Pan_2017}. 

For such applications, an alternative cavitation number was introduced by Fatjo\cite{Fatjo_2016}. It is analogous to the cavitation number $\sigma$ introduced in Eq.~\ref{eq: Ca1} but replaces the dynamic pressure in the denominator with the \textit{water-hammer pressure} obtained from rigid column theory. This yields the following formulation:
\begin{equation}
    \mathrm{Ca} = \frac{p_1-p_\mathrm{v}}{\rho_\ell\,a\,h},
    \label{eq: Ca}
\end{equation}
where $\rho_\ell\,a\,h = \Delta p$ is the pressure increase in the liquid column due to the water-hammer effect obtained from rigid column theory, with $h$ the height of the water column, and $a$ its instantaneous acceleration. Similarly to $\sigma$, cavitation is expected to occur if $\mathrm{Ca} \ll 1$. This number has been experimentally validated by some tube-arrest and drop-tube experiments\cite{Pan_2017, Chen_2004}. The gravitational constant $g$ can be introduced into Eq.~\ref{eq: Ca} to formulate the cavitation number as a ratio between the non-dimensional force from the pressure difference and the non-dimensional acceleration, which yields
\begin{equation}
    \mathrm{Ca} = \frac{(p_1-p_\mathrm{v})/(\rho_\ell\,g\,h)}{a/g}.
    \label{eq: Ca2}
\end{equation}

Although cavitation in impulsively accelerated liquid columns has been investigated experimentally\cite{Pan_2017, Chen_2004} and with numerical simulations\cite{Veilleux_2018}, previous studies have fallen short of producing a model to quantitatively reproduce the mechanism linking the dynamics of a liquid column to tension and cavitation in the liquid. In particular, the coupling between the dynamics of the liquid column and the resulting pressure field has not been satisfactorily modeled. 

The present study is motivated by the potential appearance of cavitation in liquid metals in a concept for fusion energy termed \emph{magnetized target fusion} (MTF)\cite{Laberge}. In this concept, a preformed plasma is compressed to fusion conditions using an impulsively collapsed liquid shell. If collapse is driven by pistons injecting the liquid, the strong compression wave generated at the center of the machine may reflect as a rarefaction wave, leading to a state of tension and potential cavitation in the liquid. Cavitation onset can represent a potential source of damage to the device's operation or may be beneficial in attenuating the high-pressure wave activity generated in the device and must therefore be understood and accurately predicted. 

Although liquid metals (e.g., lithium or lithium-lead eutectic) would be used in an MTF application, the present study focuses on cavitation in water. Little is known about the cavitation behavior of liquid metals, and thus water is used as a well-quantified surrogate for liquid metal in the present study. We assume here that liquid metals will behave in a way that is analogous to our results with water. Previous research by this group has investigated the pressure pulse and cavitation onset in the collapse of a piston-driven annular cavity for a similar application using water\cite{Sobral_2023}.

This paper investigates the conditions for the onset of cavitation in a piston-driven liquid column as it impulsively compresses a gas volume. The paper is structured as follows. Section II presents the experimental apparatus of a piston-driven liquid column and the associated diagnostics. Additionally, a modeling approach is presented that reproduces the experimental dynamics and treats the liquid as a spring-mass system to capture the wave activity in the column. Section III presents representative experimental results in detail, visualized using $xt$-diagrams reconstructed from videography and synchronized with the recorded pressure activity. Section IV presents the modeling results, which serve to identify the mechanism of cavitation onset. The results of the model in comparison to a comprehensive sweep of the experimental parameter space, defining regions where cavitation is or is not observed, are presented in Section V, and a concluding discussion is found in Section VI.
\vspace{-1em}

\section{Methods}
\subsection{Experimental Setup}
\subsubsection{Experimental Apparatus}

The experimental apparatus is shown schematically in Fig.~\ref{fig: Exp_Ap}. It consists of a polycarbonate tube of height $H~=~310$~mm, with an inner diameter of 38.1 mm (1.5 in) and wall thickness of 3.18 mm (1/8 in), sealed at both ends by aluminum flanges and containing an aluminum piston of length 31.3 mm (1.23 in). The piston has two lubricated rubber X-rings that provide a dynamic seal to prevent gas or water leaks around the piston. The sealed volumes above and below the piston are independently connected through plumbing to a vacuum pump. The volume below the piston is maintained at approximately 1 kPa before every experiment to ensure the piston starts at the bottom of the tube. The volume under the piston is also connected to a high-pressure gas reservoir (volume 500 mL) through a fast-acting solenoid valve (model Granzow E2B19-000). The development of this apparatus underwent several iterations, the details of which are reported in Kokkalis\cite{Kokkalis_2023}.

Prior to an experiment, distilled water is introduced in the volume above the piston to the desired height $h$, leaving part of the volume filled with gas. The water is degassed \textit{in situ} by lowering the pressure to around 1~kPa for 2--5~minutes to limit the presence of dissolved gasses and small bubbles in the water before an experiment, thus mitigating the effects of gaseous inclusions on the threshold of cavitation onset. This is repeated multiple times until no bubbles are visible in the water column. The pressure in the test section gas volume is then set to the desired value ($p_\mathrm{t,0}$) using the vacuum pump, and the section is then closed off from the vacuum line with a valve. The gas reservoir is filled with compressed nitrogen to the desired driving pressure ($p_\mathrm{r,0}$) before the start of the experiment. 

The experiment begins by triggering the solenoid valve to open, allowing nitrogen to expand through the valve and push the piston upward in the polycarbonate test section. The piston accelerates the liquid column upward and compresses the gas volume above the water. The pressure in the gas volume eventually reaches a maximum that causes the water column and piston to stop abruptly and then to start moving downward, as the gas volume expands and decompresses. This event is termed \textit{turnaround} and is analogous to a water-hammer flow, in which a liquid column is subjected to a quick deceleration leading to a large overpressure. Cavitation---if it occurs---is observed shortly after turnaround, and the experiment is over 15--30~ms after the initial opening of the solenoid valve.

Three experimental parameters can be independently controlled, determining the dynamic and compression profile of the liquid column in the test section. The first parameter is the height of water ($h$), which can be varied between two extremes: a case with no water (only gas above the piston) and a case with no gas (only water above the piston). The second parameter is the initial pressure in the gas reservoir ($p_\mathrm{r,0}$) prior to the experiment, which can be varied up to values of 20~bar. As the initial pressure in the gas reservoir is increased, the piston is driven up at greater accelerations and the liquid column reaches greater peak velocities. The third independent parameter is the initial pressure in the test section ($p_\mathrm{t,0}$) set before an experiment, which can be varied between 1--100 kPa. As this pressure is lowered, the mass of gas in the test section decreases and is therefore more easily compressed, causing the liquid column to reach higher peak velocities, accelerations, and peak test section pressures.

The three parameters can be controlled independently to change the dynamics of the experiment, resulting in experimental peak velocities in the range of 7--16~m/s, accelerations of 4\,000--90\,000~m/s$^2$, and peak transient pressures ranging from 1 to 6~MPa. The parameters $p_\mathrm{r,0}$ and $p_\mathrm{t,0}$ have no physical meaning outside of the context of this experiment, therefore we choose in this paper to report experiments quantified by the height of the liquid column $h$, the maximum piston velocity $v_\mathrm{p, max}$, and the maximum test section pressure $p_\mathrm{t, max}$.

\vspace{-1em}

\subsubsection{Diagnostic Methods}
The dynamics of the piston and water column are recorded via high-speed videography (Photron FASTCAM SA5 model 1300K-M1) with a frame rate of 50\,000~fps and a resolution of 128$\times$776~pixels (spatial resolution of 4~mm/pixel). The position of the piston in time is tracked using an in-house MATLAB image-tracking code, which uses a Kanade-Lucas-Tomasi point-tracking function from the MATLAB Computer Vision Toolbox. The position of the piston is extracted at each frame and velocity and acceleration are obtained via central-difference schemes.

The high-speed imagery also allows us to see the bubble cluster growth and collapse in the liquid column. In order to visualize a complete experiment in time as a single plot, a post-processing method was written in MATLAB in which each frame is averaged horizontally into a single pixel column. The individual pixel column for each frame is then displayed stacked side by side with the other pixel columns to provide an $xt$-diagram of the experiment with time on the horizontal axis and height on the vertical axis. This allows us to visualize the dynamics of an experiment as well as the location and timescales of bubble growth and collapse in a single plot.

In addition to high-speed imagery, pressure measurements are used to record the acoustics in the test section. The pressure in the gas volume of the test section is measured with a PCB-113B23 (PCB Piezotronics) pressure sensor connected to a digital oscilloscope (PicoScope model~4824) sampling at 1~MHz with a resolution of 3.45~kPa/mV. This measurement allows us to quantify the compression rate and amplitude of the gas volume in the test section that link to the dynamics of the experiment and the onset of cavitation. 

To obtain pressure measurements directly in the liquid column, a modified polycarbonate tube can be used that has two additional pressure sensors along its length. The sensors are mounted horizontally into two custom-built aluminum flanges such that the sensor face is flush with the inside of the tube. The two sensors (one PCB-113A26 with a resolution of 68.8~kPa/mV and one PCB-113B23 with a resolution of 3.43~kPa/mV) are mounted at a height of 125~mm and 213~mm from the bottom of the tube, respectively. The attachment feature for these sensors partially obstructs the view of the test section, and the water height must be at least $h=200$~mm to prevent the piston from passing in front of the bottom sensor. Therefore, these two additional sensors are only used for a small subset of experiments to give insight into the pressure field in the liquid column and to validate the model results.
\begin{figure}[b!]
\includegraphics[width = \linewidth]{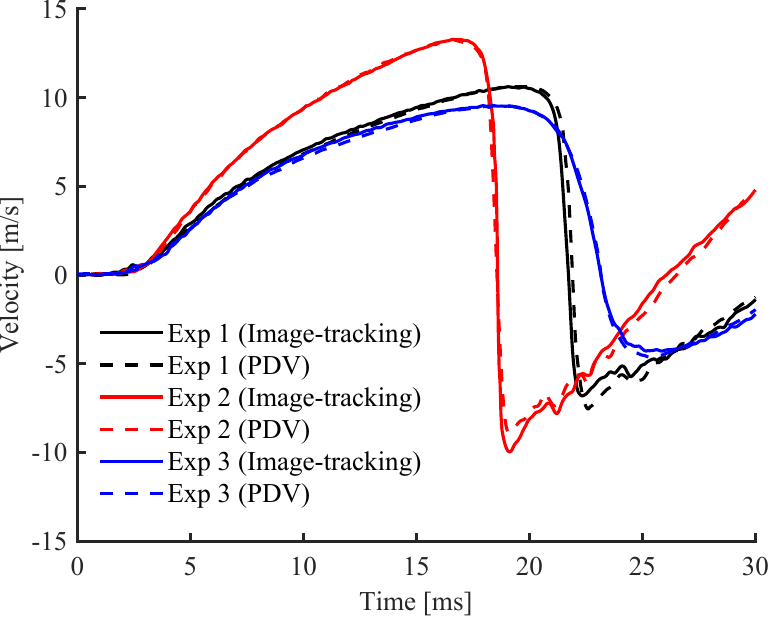}% Here is how to import EPS art
\caption{\label{fig: PDV_comp} Typical velocity traces obtained from the PDV measurements and the image-tracking method. Experimental parameters for Exp~$(1,\,2,\,3)$ are, respectively: $h = (145,\,145,\,145)$~mm, $p_\mathrm{t, max} = (3.9,\,4.6,\,1.5)$~MPa, $v_\mathrm{p, max} = (10.3,\,13.0,\,9.35)$~m/s. }
\end{figure}
\vspace{-1em}
\subsubsection{Validation of Image-Tracking}

The piston dynamics extracted from the high-speed imagery were validated against data from a Photonic Doppler Velocimetry (PDV) system\cite{Strand_2006}. PDV measures the piston velocity directly and accurately by reflecting a laser beam off the piston's bottom surface and measuring the variation in the beat frequency obtained by blending the Doppler-shifted reflected beam and a reference beam\cite{Dolan}. The PDV probe in the present experiment is mounted on the bottom flange of the apparatus, and the laser beam is directed through a small transparent window that allows good optical access to the bottom surface of the piston. 

Figure~\ref{fig: PDV_comp} shows a comparison between velocity profiles obtained with PDV and high-speed video for typical experiments. We can see that there is good agreement between the two methods, thus validating the accuracy of the piston-tracking method to obtain velocity. Additionally, we find that the acceleration values obtained using the second derivative of video-tracking measurements fall within 10\%, on average, of the acceleration obtained from the first derivative of PDV velocity data. Due to the relative complexity of the PDV setup and the good agreement with high-speed camera measurements, the experimental data shown hereafter is obtained from the image-tracking method rather than PDV measurements.
\vspace{-1em}

\subsection{Modeling Framework and Derivations}
\label{sec: Model_Dvlpmt}
The model introduced in this paper represents a novel approach to treating flows of liquid columns undergoing acceleration. This approach is purely one-dimensional and aims to couple dynamics modeling to the liquid column acoustics that are of key importance for understanding and predicting cavitation. The equations presented herein serve to create a numerical tool that can simulate the experimental setup given a set of parameters that define the experiment geometry and its initial conditions. 

The modeling approach uses the following simplifying assumptions to model the dynamics. The compression and expansion of the gas in the test section, as well as the gas discharge from the reservoir, are assumed to be isentropic. This is because the experiment happens quickly enough that the gas behaves adiabatically. The friction of the piston, as well as the friction due to the water on the walls, were found to have a negligible effect compared to the large pressure forces acting on the system and are therefore ignored. The model is one-dimensional and therefore does not capture the surface instabilities of the liquid. The surface instabilities are not believed to significantly affect the pressure field in the liquid due to the small magnitude of the instabilities compared to the large pressure magnitude in the test section (order of 1--10~MPa). Finally, the valve is assumed to behave like a simple orifice that opens instantaneously. Calibration experiments of gas discharge through the valve orifice were conducted and showed that the valve opening time had a negligible effect on the pressure discharge behavior and on the dynamics of the experiment. Additionally, these experiments were used to obtain an effective orifice area for the valve ($A_\mathrm{v}$) which includes losses through the valve (equivalent to determining a discharge coefficient).
\vspace{-1em}

\subsubsection{Gas Reservoir Discharge Modeling}
The model aims to accurately reproduce the dynamics of the experiment while also providing a solution for the pressure field in the liquid column. To do that, the model must take into account the transient discharge of the gas contained in the volume of the reservoir ($V_\mathrm{r}$) into the volume under the piston ($V_\mathrm{p}$). This is done by modeling the flow through the solenoid valve as compressible gas flow through an orifice, subject to a condition of choking if $p_\mathrm{p} < 0.5283p_\mathrm{r}$. The flow rate out of the reservoir is given by
\begin{equation}
     \frac{\mathrm{d}m}{\mathrm{d}t} = -A_\mathrm{v}p_\mathrm{r}\sqrt{\frac{\gamma}{RT_\mathrm{r}}}\left(\frac{2}{\gamma+1}\right)^{\frac{\gamma+1}{2(\gamma-1)}}
     \label{eq: dmdt1}
 \end{equation}
 if the orifice is choked and is otherwise given by
 \begin{equation}
     \frac{\mathrm{d}m}{\mathrm{d}t} = - \frac{A_\mathrm{v}p_\mathrm{r}}{\sqrt{RT_\mathrm{r}}} \left( \frac{p_\mathrm{p}}{p_\mathrm{r}} \right)^{\frac{1}{\gamma}} \left[ \frac{2\gamma}{\gamma - 1}  \left( 1 - \left( \frac{p_\mathrm{p}}{p_\mathrm{r}} \right)^{\frac{\gamma - 1}{\gamma}} \right) \right]^{\frac{1}{2}},
     \label{eq: dmdt2}
 \end{equation}
 where $A_\mathrm{v}$ is the effective orifice area of the valve, $R$ is the specific gas constant of nitrogen, $T_\mathrm{r}$ is the temperature of the gas in the reservoir (obtained using the ideal gas law), and $\gamma$ is the ratio of specific heats for nitrogen. 

Knowing the flow rate, the equation for the rate of change of pressure in the reservoir is easily derived from the ideal gas law using isentropic relations and taking into account that the volume stays constant:
 \begin{equation}
     \frac{\mathrm{d}p_\mathrm{r}}{\mathrm{d}t} = \frac{\gamma RT_\mathrm{r}}{V_\mathrm{r}}\frac{\mathrm{d}m_\mathrm{r}}{\mathrm{d}t},
      \label{eq: dPr_dt}
 \end{equation}
where $m_\mathrm{r}$ is the mass of gas in the reservoir volume. The volume below the piston is subject to a mass flow rate in and a volume $V_\mathrm{p}$ that changes as the piston moves up and down in the column. A conservation of energy approach\cite{Richer_2000} is used to model the pressure under the piston as a function of the mass flow rate and the rate of change of volume
\begin{equation}
     \frac{\mathrm{d}p_\mathrm{p}}{\mathrm{d}t} = \frac{\gamma-1}{V_\mathrm{p}}c_{p}\frac{\mathrm{d}m_\mathrm{p}}{\mathrm{d}t}T_\mathrm{r}-\gamma \frac{p_\mathrm{p}}{V_p}\frac{\mathrm{d}V_\mathrm{p}}{\mathrm{d}t},
 \end{equation}
 where $c_{p}$ is the heat capacity of nitrogen at constant pressure, and $m_\mathrm{p}$ and $p_\mathrm{p}$ are the mass and pressure of nitrogen in the volume under the piston, respectively. We can use the relationship between the volume under the piston and the piston position $x_\mathrm{p}$ to obtain
\begin{equation}
     \frac{\mathrm{d}p_\mathrm{p}}{\mathrm{d}t} = \frac{1}{V_\mathrm{p}}\left((\gamma-1)c_{p}\frac{\mathrm{d}m_\mathrm{p}}{\mathrm{d}t}T_\mathrm{r}-\gamma p_\mathrm{p}A_\mathrm{p}\frac{\mathrm{d}x_\mathrm{p}}{\mathrm{d}t}\right),
      \label{eq: dPp_dt}
 \end{equation}
 where $A_\mathrm{p}$ is the area of the bottom face of the piston that is exposed to pressure. 

\subsubsection{Acoustic Modeling Approach and Wave Speed}
\begin{figure}[t!]
\includegraphics[width = 0.35\textwidth]{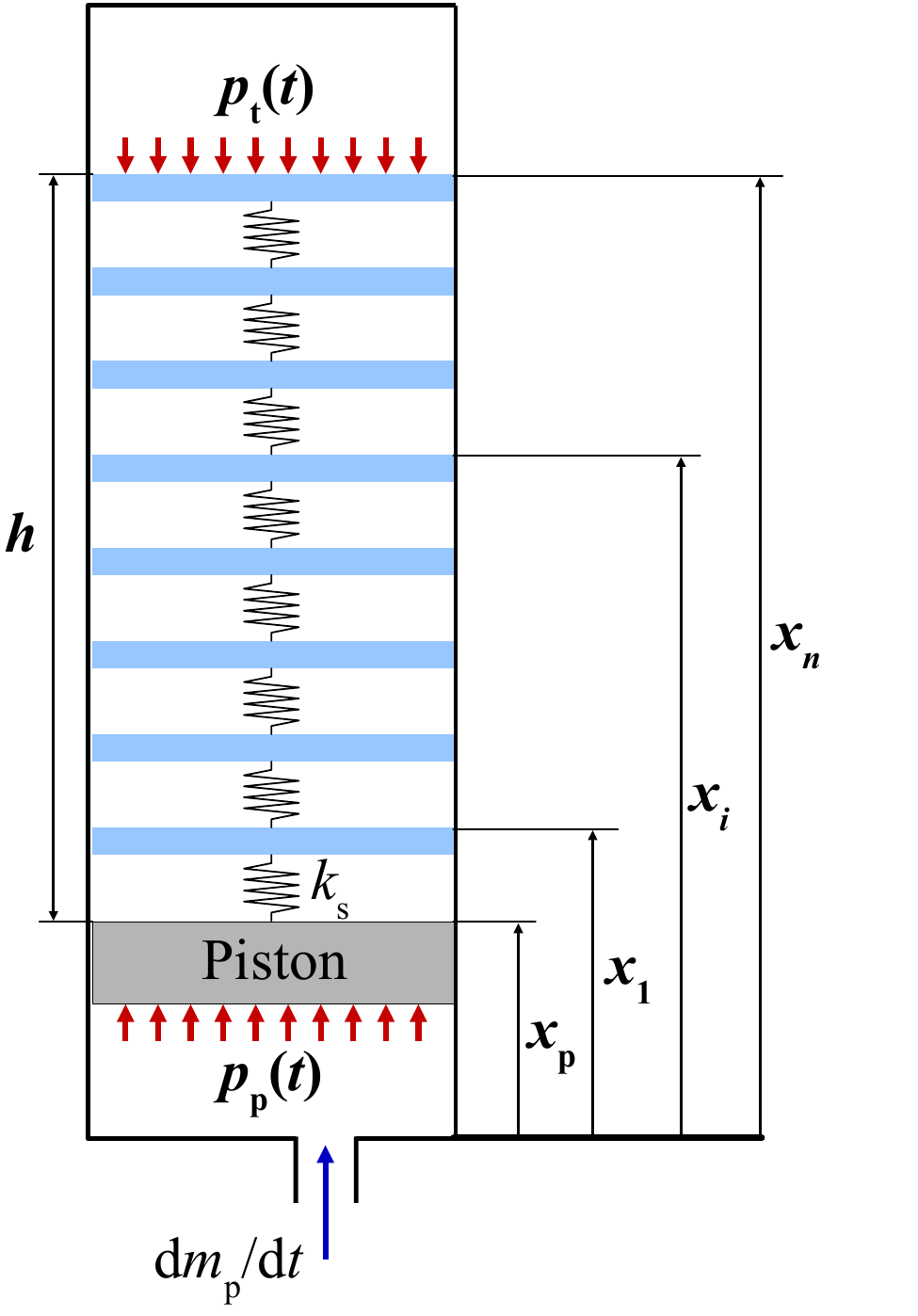}% Here is how to import EPS art
\caption{\label{fig: Model_Framework} Schematic representation of the model framework with the liquid column discretized into spring-connected fluid elements. The number $n$ of fluid elements shown here is arbitrary.}
\end{figure}

To correctly model the acoustics in the liquid column, a novel approach is introduced in which the compressibility of the water is taken into account by modeling it as a system of spring-connected masses. This allows us to capture the coupling between the dynamics of the water column and the acoustic waves traveling through it. Figure~\ref{fig: Model_Framework} shows a schematic representation of the model framework. The water column of height $h$ is divided into $n$ fluid elements, each of length $L_0=h/n$. Each element has mass $m_\mathrm{e} = m_\mathrm{w}/n$, where $m_\mathrm{w}$ is the total mass of the liquid column. 

We can relate the stiffness of the springs in the model to the bulk modulus of the liquid to ensure that the speed of sound in the coupled spring-mass system reproduces the correct speed of sound. The bulk modulus ($\kappa$) is defined as the change in pressure in the liquid divided by the volumetric strain, which can be expressed as
\begin{equation}
    \kappa = \frac{\Delta p}{\Delta V / V_0},
    \label{eq: K1}
\end{equation}
 where $\Delta p$ is the change in pressure, $\Delta V$ is the change in volume, and $V_0$ is the original volume of fluid. Using a conservation of mass argument, the expression above can be modified to obtain the well-known equation for bulk modulus
\begin{equation}
    \kappa = \rho_\ell \,c^2,
    \label{eq: K2}
\end{equation}
where $c$ is the wave speed of the fluid. 

In our one-dimensional system, the volume of one fluid element is equal to its original length $L_0$ times the cross-sectional area of the inside of the tube $A_\mathrm{p}$. The volume change can therefore be expressed as $\Delta V = A_\mathrm{p}\Delta x$, where $\Delta x$ represents the change in height of the fluid element due to compression. We can also express the change in pressure as the force acting on a spring over the area $A_\mathrm{p}$, such that $\Delta p = \Delta F / A_\mathrm{p}$. Using these relations and equating the right-hand sides of Eqs.~\ref{eq: K1} and \ref{eq: K2}, we obtain:
\begin{equation}
    \frac{L_0}{A_0}\frac{\Delta F}{\Delta x} = \rho_\ell \,c^2.
\end{equation}
Noticing that the spring stiffness is given as $k_\mathrm{s} = \Delta F / \Delta x$, and that the density can be expressed as $\rho_\ell = m_\mathrm{e}/V_0$, we can obtain an expression for the spring stiffness that must be used in the model
\begin{equation}
    k_\mathrm{s} = \frac{m_\mathrm{e}\,c^2}{L_0^2}.
    \label{eq: k_s}
\end{equation}

Because the liquid in our experiment is contained in a thin-walled polycarbonate tube, we must consider the effects of fluid-solid coupling on the wave speed. Korteweg theory predicts that the fluid-structure interaction decreases the speed at which waves propagate in the system\cite{Shepherd_2010}. The amount of fluid-solid coupling can be quantified by the following parameter: $\beta = 2\kappa r / Ez$, where $r$ is the radius of the tube, $E$ is the elastic modulus of the tube material, and $z$ is the thickness of the tube wall\cite{Shepherd_2010}. In our experiment, using $E = 2.4$~GPa for polycarbonate, $z = 3.175$~mm, $r = 19.05$~mm, and $\kappa=2.2$~GPa for water, we obtain a coupling parameter $\beta = 11$. Since $\beta \gg 1$, acoustic waves in the tube are expected to propagate at the Korteweg wave speed\cite{Shepherd_2010}, which is given by
\begin{equation}
    c = \frac{a_\mathrm{f}}{\sqrt{1+\beta}},
\end{equation}
where $a_\mathrm{f}$ is the speed of sound in the fluid. For water, we may take $a_\mathrm{f}\,=\,1488$~m/s, and thus obtain a Korteweg wave speed of $c\,=\,430$~m/s, which is the wave speed we use in Eq.~\ref{eq: k_s}.

With the formulation of liquid compressibility as a system of spring-connected masses, we can now model the acoustics in the liquid column as it undergoes the dynamics of the experiment. This can be used to investigate the mechanism by which some springs may come under tension, thus indicating a potential site of cavitation.

\subsubsection{Model Formulation}
We now have established an approach to treat the discharge of the gas reservoir into the volume under the piston, as well as a way to treat the water as a compressible fluid to capture its dynamics and the acoustics through the liquid column. This is now formulated as a system of second-order non-linear coupled ODEs that is solved numerically using the built-in MATLAB Runge-Kutta solver. The state vector is given by
\begin{align}
    y &= \begin{bmatrix}
           m_\mathrm{r} \\
           p_\mathrm{r} \\
           m_\mathrm{p} \\
           p_\mathrm{p} \\
           x_\mathrm{p} \\
           \Dot{x}_\mathrm{p}\\
           x_1\\
           \Dot{x}_1\\
           ...\\
           x_n\\
           \Dot{x}_n\\
         \end{bmatrix}.
  \end{align}
The variables $m_\mathrm{r}$ and $m_\mathrm{p}$ correspond to the gas mass, and $p_\mathrm{r}$ and $p_\mathrm{p}$ correspond to the gas pressures in the reservoir and under the piston, respectively. Variables $x_\mathrm{p}$ and $\Dot{x}_\mathrm{p}$ correspond to the position and velocity of the piston, and $x_{i}$ and $\Dot{x}_{i}$ correspond to the position and velocity of each fluid element for $i=1,2,\ldots,n$. 

We now take the derivative in time of the state vector to obtain
\begin{align}
    \Dot{y} &= \begin{bmatrix}
           \Dot{m}_\mathrm{r}\\
           \Dot{P}_\mathrm{r}\\
           \Dot{m}_\mathrm{p}\\
           \Dot{P}_\mathrm{p}\\
           \Dot{x}_\mathrm{p}\\
           \Ddot{x}_\mathrm{p}\\
           \Dot{x}_1\\
           \Ddot{x}_1\\
           ...\\
           \Dot{x}_n\\
           \Ddot{x}_n\\
         \end{bmatrix}.
  \end{align}
The mass flow rates $\Dot{m}_\mathrm{r}$ and $\Dot{m}_\mathrm{p}$ are equal and opposite to each other, and they are evaluated using Eqs.~\ref{eq: dmdt1} or \ref{eq: dmdt2}, subject to the condition of choked or unchoked flow. The pressures $p_\mathrm{p}$ and $p_\mathrm{r}$ are solved using Eq.~\ref{eq: dPp_dt} and Eq.~\ref{eq: dPr_dt}, respectively. The equations for velocity are expressed as $\Dot{x}_i = \mathrm{d}x_i/\mathrm{d}t$ for $i=1,2,\ldots,n$. 

The equations for acceleration are obtained for the piston and for each fluid element using Newton's second law. The piston is pushed from below by the pressure $p_\mathrm{p}$, and from above by the spring of the first fluid element. We can therefore express its equation of motion as
\begin{equation}
    \Ddot{x}_\mathrm{p} = \left(p_\mathrm{p}(t)A_\mathrm{p} + k_\mathrm{s}(x_1 - x_\mathrm{p} - L_0)- m_\mathrm{piston}\,g\right)/m_\mathrm{piston},
\end{equation}
where $m_\mathrm{piston}$ is the mass of the piston. For fluid elements $i=1,2,\ldots,(n-1)$, the equation of motion is formulated as
\begin{equation}
    \Ddot{x}_{i} = \left(-k_\mathrm{s}(x_{i} - x_{i-1} - L_0) + k_\mathrm{s}(x_{i+1} - x_{i} - L_0) - m_\mathrm{e}\,g\right)/m_\mathrm{e}.
\end{equation}
Finally, the equation of motion for the last fluid element is
\begin{equation}
    \Ddot{x}_{n} = \left(-p_\mathrm{t}(t)A_\mathrm{p} - k_\mathrm{s}(x_{n} - x_{n-1} - L_0)- m_\mathrm{e}\,g\right)/m_\mathrm{e},
\end{equation}
where $p_\mathrm{t}(t)$ is obtained assuming the gas volume in the test section undergoes isentropic compression where $PV^\gamma$ remains constant. We have formulated coupled equations for all variables in the state vector, allowing us to solve it numerically given a set of initial conditions. These equations can be solved numerically in only a few seconds, thus allowing us to run many cases over a wide range of parameters.

\section{Experimental Results}
\subsection{Cavitation Onset and Collapse Behavior}
\label{sec: Cavi_Onset}
\begin{figure*}[]
        \includegraphics[width = \textwidth]{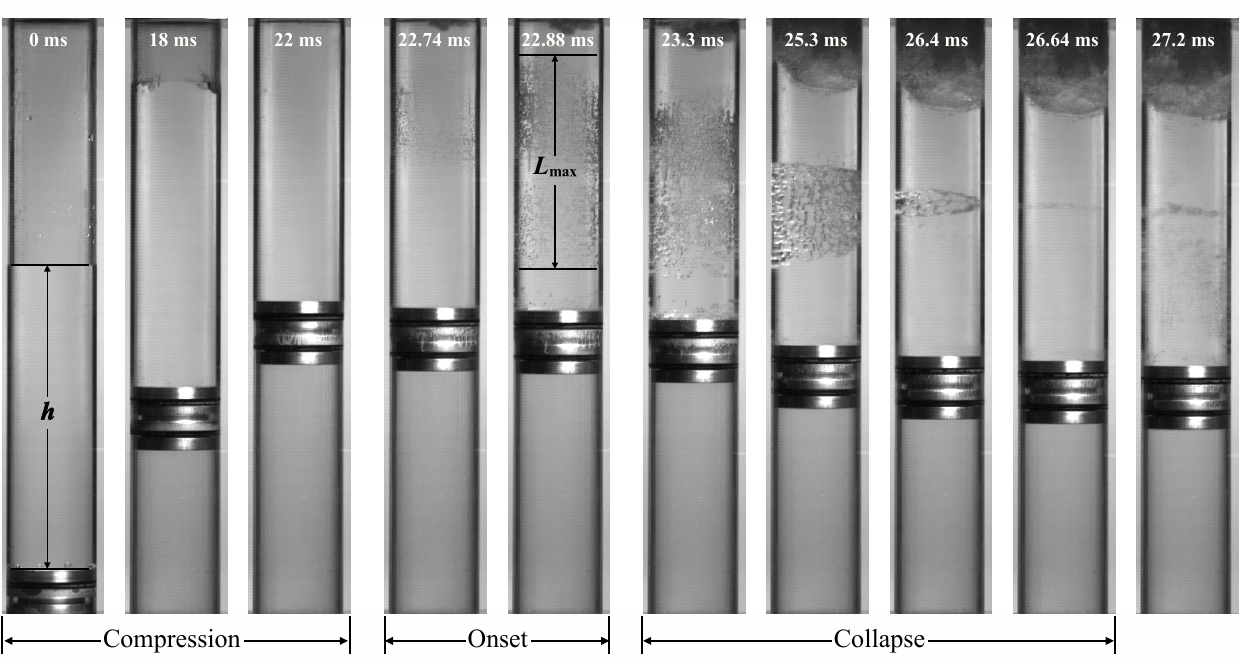}
\caption{\label{fig: Exp66_Images} High-speed images of a typical experiment with cavitation cluster formed in the liquid column (multimedia available online). The experimental parameters for the experiment shown here are: $h = 150$ mm, $p_\mathrm{t,max} = 4.0$~MPa, and $v_\mathrm{p,max} = 10.5$~m/s.}
\end{figure*}
\begin{figure}[b!]
\includegraphics[width = 0.5\textwidth]{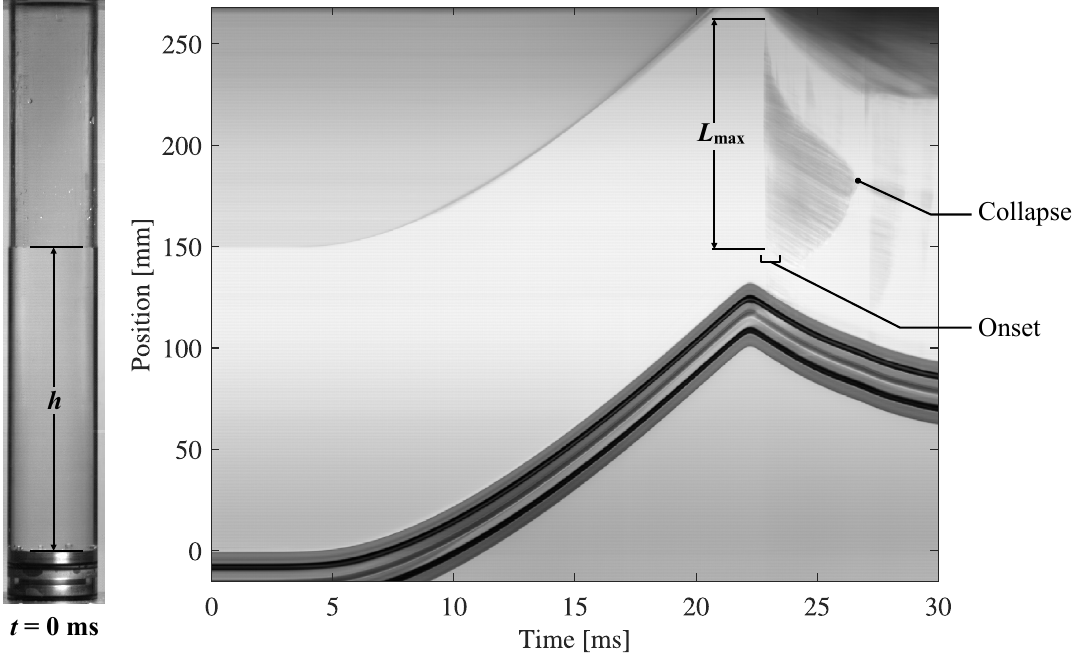}% Here is how to import EPS art
\caption{\label{fig: Exp66_xt_diagram} Experimental $xt$-diagram obtained for an experiment with parameters: $h = 150$ mm, $p_\mathrm{t,max} = 4.0$~MPa, and $v_\mathrm{p,max} = 10.5$~m/s.}
\end{figure}

We first investigate the onset of cavitation for one representative experiment. Figure~\ref{fig: Exp66_Images} shows high-speed images from a typical experiment where cavitation occurs in the liquid column (multimedia available online). This experiment has a liquid column height of 150~mm, reaches a peak pressure of 4.0~MPa, and a peak velocity of 10.5~m/s. The experiment is broken down into three distinct phases: the compression of the gas volume, the onset of cavitation, and the collapse of the bubble cluster. The compression phase occurs when the piston and water column move upward, compressing the gas volume above the water, and ends at the moment of turnaround. This can take anywhere between 15--30~ms depending on the experimental parameters (22~ms in the experiment shown in Fig.~\ref{fig: Exp66_Images}). During this phase, upward velocities typically reach 10 to 15~m/s, with accelerations up to 7\,000~m/s², and no cavitation is observed. 

\begin{figure*}[]
        \includegraphics[width = \textwidth]{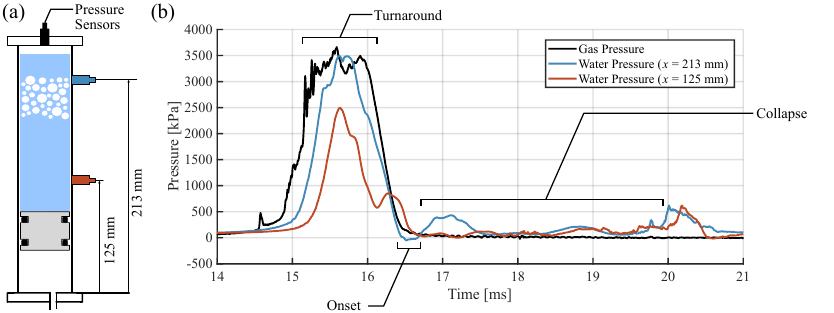}
\caption{\label{fig: Pressure_Field} (a) Schematic representation of the location of the three pressure measurements during an experiment. (b) Pressure measurements from the top of the test section and two locations along its height (125~mm and 213~mm) for experimental parameters: $h = 200$ mm, $p_\mathrm{t,max} = 3.5$~MPa, and $v_\mathrm{p,max} = 8.33$~m/s. The pressure data has been smoothed using a moving-average filter.}
\end{figure*}

The moment of turnaround corresponds to when the gas in the test section reaches a maximum pressure and a minimum volume and causes the liquid column to stop abruptly. This is similar to a water-hammer flow, however, in our experiment, the small gas volume provides some cushioning of the liquid column impact. This also corresponds with the highest acceleration of the liquid column, with values of up to 90\,000~m/s$^2$. In our experiments, near turnaround, the free surface of the liquid is no longer visible as the small gas volume is hidden behind the aluminum flange (see Fig.~\ref{fig: Exp66_Images} and \ref{fig: Exp66_xt_diagram}). Immediately after turnaround, the liquid column starts accelerating downward due to the high pressure in the test section. As the piston starts moving downward, a state of tension may be reached in the liquid column that leads to cavitation onset. 

Cavitation is identified visually by the initial appearance and sudden growth of cavitation bubbles in the test section. The initial onset of cavitation happens between 0.4--1 ms after turnaround, depending on the experiment. This is followed by a short and intense period of bubble growth lasting about 0.1~ms, after which the bubble cluster reaches its maximum size. We identify this point as the state of maximum cavitation in the liquid column and quantify the amount of cavitation by measuring the height $L_\mathrm{max}$ of the bubble cluster. Figure~\ref{fig: Exp66_xt_diagram} shows the $xt$-diagram for the experiment, in which we can clearly visualize the motion of the piston and liquid column as well as the bubble cluster that appears as a darker region shortly after turnaround.

\begin{figure}[b!]
\includegraphics[width = 0.45\textwidth]{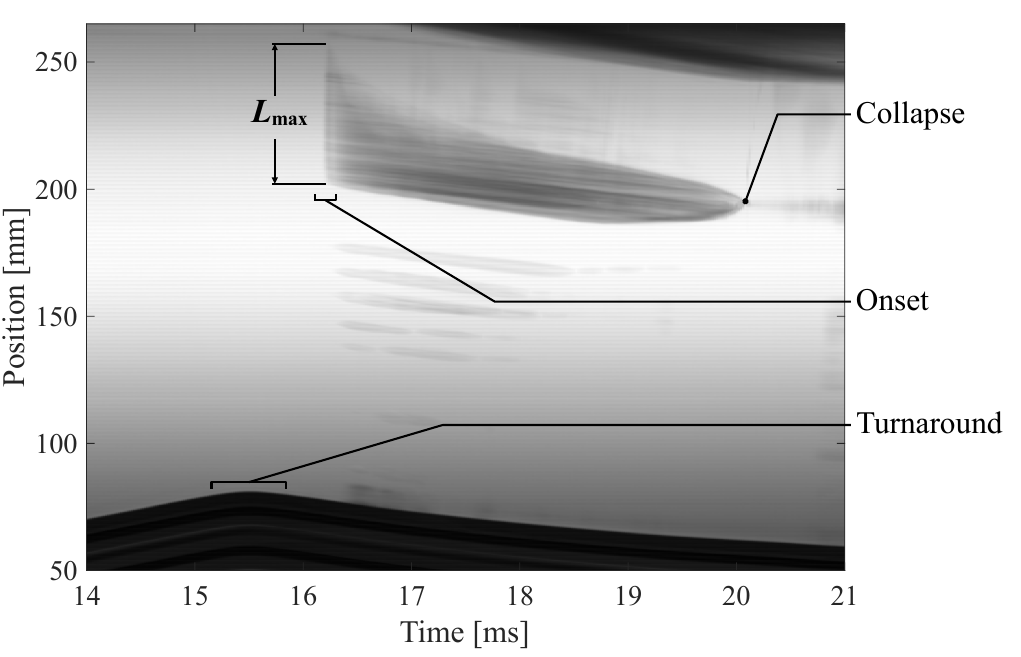}% Here is how to import EPS art
\caption{\label{fig: Ex192_xt_diagram} Experimental $xt$-diagram obtained for an experiment with parameters: $h = 200$ mm, $p_\mathrm{t,max} = 3.5$~MPa, and $v_\mathrm{p,max} = 8.33$~m/s.}
\end{figure}

The third phase of the experiment occurs after the initial onset of cavitation and corresponds to the collapse of the bubble cluster. This phase can last between 0.01--5~ms depending on the experimental conditions (3.34~ms in the experiment shown in Fig.~\ref{fig: Exp66_Images}). During the collapse phase, individual bubbles may continue to grow in size, but the bubble cluster size continuously decreases, as seen in Fig.~\ref{fig: Exp66_xt_diagram}. As the bubble cluster size decreases, individual bubbles merge to form a tightly packed group of large bubbles that all converge and collapse at a single point. After the collapse of this cluster, we observe a secondary, weaker, onset of cavitation that can be seen in the last image of Fig.~\ref{fig: Exp66_Images} and between 27 and 30~ms in Fig.~\ref{fig: Exp66_xt_diagram}.

The onset and collapse behavior shown in Fig.~\ref{fig: Exp66_Images} and \ref{fig: Exp66_xt_diagram} is qualitatively similar for all experiments where cavitation is seen. The location and size of the bubble cluster, as well as the timing of turnaround, cavitation onset, and collapse are different, however, and depend on the specific dynamics of a given experiment. A key observation from these results is that the location of cavitation is generally not at either end of the liquid column, but rather occurs at some location along its length, with little to no cavitation seen near the piston or the free surface. Additional experimental results of cavitation can be found in Appendix~\ref{App: Exp}.

\subsection{Pressure Field in the Liquid Column}
\label{sec: Pressure_Field_Exp}
We now investigate the pressure field in the liquid column during and after turnaround as it relates to the mechanism and location of cavitation onset. Figures~\ref{fig: Pressure_Field} and \ref{fig: Ex192_xt_diagram} show pressure measurements and an $xt$-diagram from an experiment where the liquid column height is 200~mm. The pressure measurements are taken in the gas volume at the top of the test section and in the liquid column at two locations along the polycarbonate tube. These two figures allow us to link the pressure field in the liquid column with the key phases of cavitation onset and collapse. 

The compression phase for this experiment lasts until turnaround at 15.5~ms, identified as a maximum in pressure of about 3.5~MPa in Fig.~\ref{fig: Pressure_Field} and a maximum in piston position in Fig.~\ref{fig: Ex192_xt_diagram}. Though the compression phase lasts 15.5~ms, most of the increase in pressure in the gas volume occurs within one millisecond before turnaround, in a way that is similar to the pressure increase in a water-hammer flow. The pressure in the liquid column decreases as fast as it increased, less than one millisecond after turnaround. We also note that the pressure measured closest to the piston (sensor at 125~mm) is about 1~MPa lower than that observed near the top.
\begin{figure*}[]
\includegraphics[width = \textwidth]{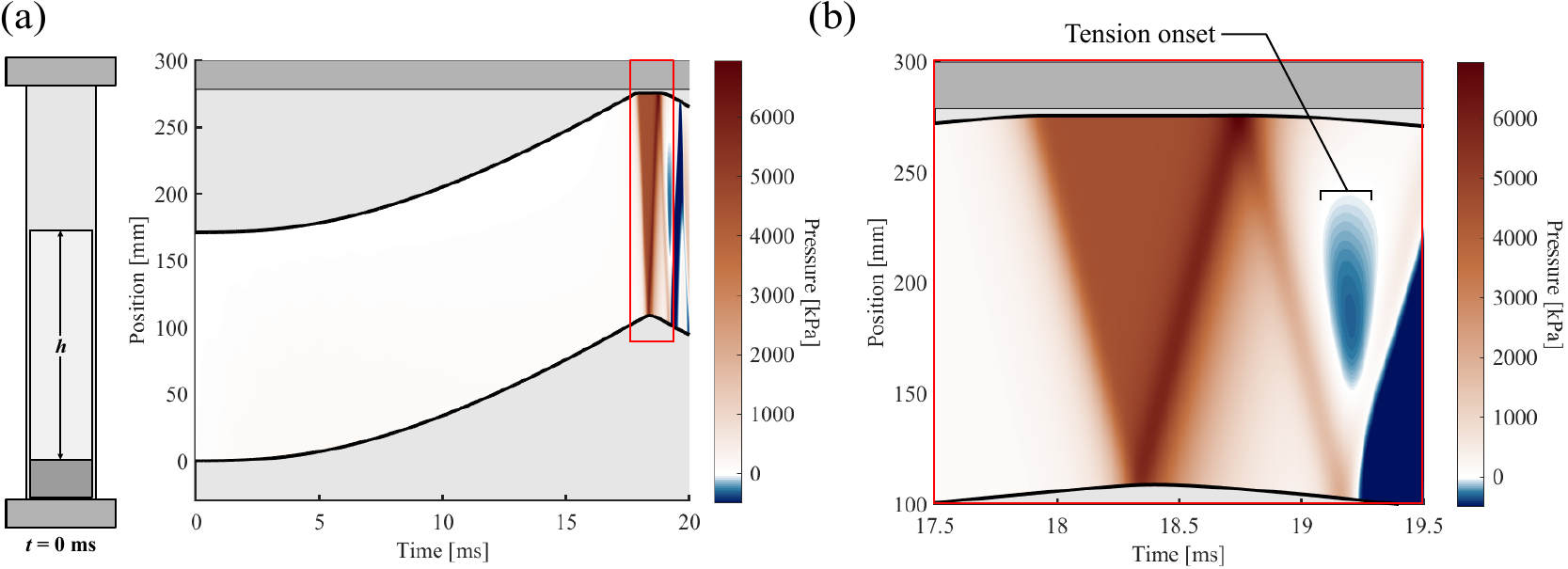}% Here is how to import EPS art
\caption{\label{fig: Model_Pressure} Modeled $xt$-diagram and pressure contours in water for a simulation where $h = 175$~mm, $p_\mathrm{t,max} = 7.1$~MPa, and $v_\mathrm{p,max} = 10.3$~m/s. (a) Full $xt$-diagram with a schematic representation of the experimental geometry. (b) Closeup of the pressure contours in water near turnaround.}
\end{figure*}

The $xt$-diagram allows us to identify the onset of cavitation of the bubble cluster less than 1~ms after turnaround at a vertical position between 200 and 260~mm along the height of the tube. At the same time, the sensor measuring pressure in the water at a position of 213~mm shows a drop in pressure below zero. The observed negative pressure matches both the time and location of the cavitation onset seen in the $xt$-diagram. We also notice that the sensor positioned at 125~mm, although it does measure a pressure drop, does not measure any pressure below zero. This again agrees with the observation that there is no cavitation observed at this position in the $xt$-diagram.

In addition to measuring the pressure drop that is simultaneous with cavitation onset, both sensors in the liquid column measure a sharp increase in pressure at about 20~ms, immediately after the end of collapse. This strong pressure wave activity corresponds to the pressure wave generated by the end of the collapse seen at about 20~ms in the $xt$-diagram of Fig.~\ref{fig: Ex192_xt_diagram}. Using pressure measurements from the submerged sensors that are at a known distance from each other, we can also measure the time delay of a passing wave to get an experimental value for the speed of sound. We obtain a value of $c = 436$~m/s, which is within 2\% of the predicted value from Korteweg's fluid-structure interaction approach.
\vspace{-1em}

\section{Modeling Results}
\label{sec: Pressure_Field_Mod}
We now investigate results obtained from a representative simulated experiment using the model developed in Section~\ref{sec: Model_Dvlpmt}. This will give us insights into how the model is able to capture the pressure field generated from the dynamics of the piston and liquid column during and after compression.
%\subsection{Pressure Field and Tension}

Figure~\ref{fig: Model_Pressure} shows a visualization of a typical modeling result that is analogous to the experimental $xt$-diagram shown in Fig.~\ref{fig: Exp66_xt_diagram}. The plot in Fig.~\ref{fig: Model_Pressure}a shows both the displacement of the piston and liquid column as well as a contour of pressure inside the water column as a function of time. The pressure at each discrete location is obtained from the force in the spring at that location divided by the cross-sectional area of the inside of the tube. The plot in Fig.~\ref{fig: Model_Pressure}b shows a closeup of the pressure contour near the moment of turnaround. The column undergoes strong compression at turnaround due to the water-hammer-like effect of compressing the gas in the test section. We observe that the volume of gas above the water surface is reduced to a height of just a few millimeters at this moment of maximum compression. The compression is closely followed by a drop in pressure as the gas volume expands and the liquid column turns around. Shortly after turnaround, a region of tension (negative pressure shown in blue) is seen in the middle of the liquid column as the piston is moving downward. 
\begin{figure}[b!]
\includegraphics[width = 0.5\textwidth]{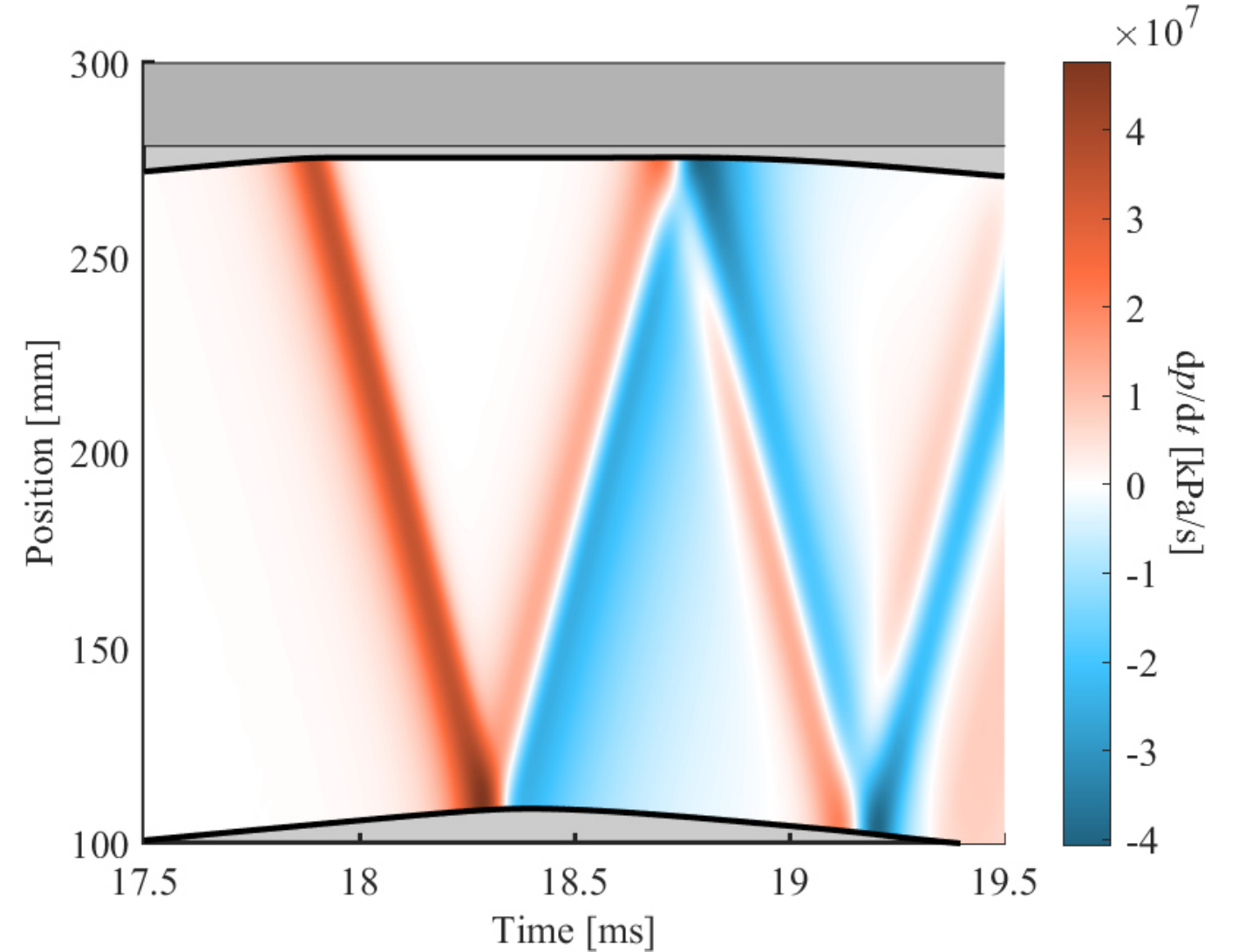}% Here is how to import EPS art
\caption{\label{fig: Model_delP} Closeup of the pressure gradient contours in water obtained with the model for a case where $h = 175$~mm, $p_\mathrm{t,max} = 7.1$~MPa, and $v_\mathrm{p,max} = 10.3$~m/s.}
\end{figure}

To better understand how the compression dynamics eventually lead to a state of tension, we take the gradient of pressure in time using a forward difference scheme, which allows us to visualize wave dynamics in the liquid column. Since the discrete mass elements used in the model are effectively Lagrangian elements, the derivative of pressure taken in time corresponds to the material derivative of pressure. Figure~\ref{fig: Model_delP} shows an $xt$-diagram visualization of this pressure gradient, allowing us to visualize wave dynamics in the liquid column. This allows us to see wave dynamics near turnaround for the same simulated experiment as shown in Fig.~\ref{fig: Model_Pressure}. 

Figures \ref{fig: Model_Pressure} and \ref{fig: Model_delP} show that for this simulated experiment, turnaround occurs at about 18.3~ms, and the initial onset of tension occurs approximately 0.7~ms after turnaround. Figure~\ref{fig: Model_delP} shows that the quick pressurization of the test section gas sends a strong compressive pulse (shown in red) downward through the liquid column which eventually reaches the piston. The earlier part of this compressive pulse reflects off of the piston first as a compressive pulse and the later part as a rarefaction pulse (shown in blue). This reflected compression wave leads to the liquid reaching a maximum pressure seen at around 18.5~ms in Fig.~\ref{fig: Model_Pressure}.

These compression and rarefaction waves lead to a successive increase and decrease in the pressure, which can be observed in Fig.~\ref{fig: Model_Pressure}. This train of compression/tension then gets reflected off of the top of the liquid column, leading to a second successive compression and tension wave going downward in the liquid column. The onset of tension seen in Fig.~\ref{fig: Model_Pressure} occurs immediately after this second rarefaction wave that is traveling downward in the liquid column.

The modeling results shown in this section for a representative experiment are qualitatively equivalent for all simulated experiments. Other simulations will differ only in the magnitude of the pressure, velocities, and accelerations for given initial conditions, as well as the timing, magnitude, and location of tension (if any) in the liquid column. We also note that the results shown in this section are comparable to those obtained using a method of characteristics (MOC) approach to track wave dynamics. The novelty of our modeling approach is the coupling of the motion and pressure field of the liquid column, which removes the need to treat complicated boundary conditions that are often a limiting factor in MOC codes. We note, however, that the model does not capture bubble growth and collapse mechanisms, and the pressure field is therefore only representative of the experiment up to the moment of cavitation onset.
\begin{figure*}[t!]
        \includegraphics[width = \textwidth]{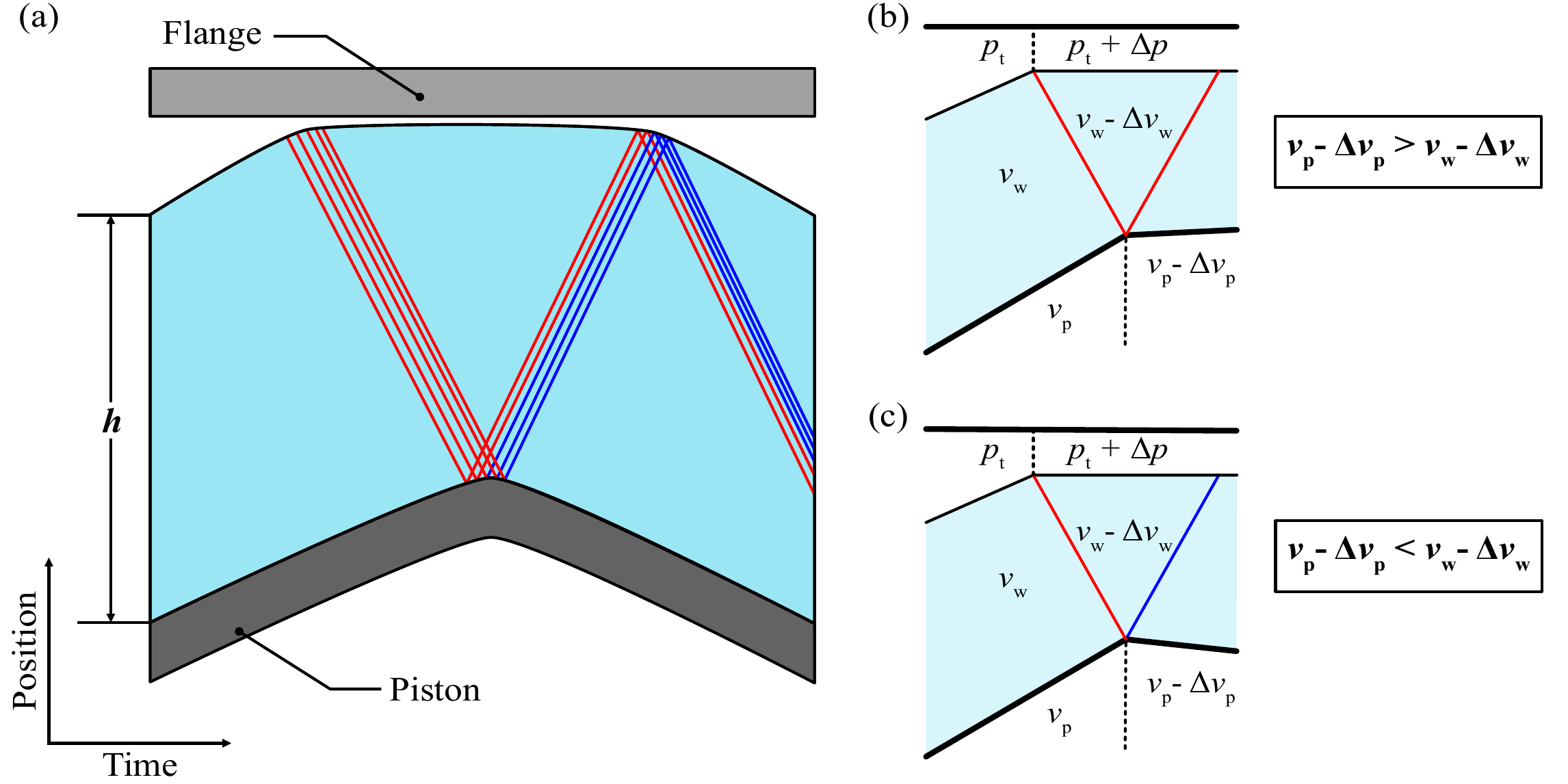}
\caption{\label{fig: Pressure_Reflections} (a) Schematic $xt$-diagram showing the generation and reflection of compression/rarefaction (red/blue) waves in the liquid column at turnaround. (b) Schematic representation of a compression--compression type reflection. (c) Schematic representation of a compression--rarefaction type reflection.}
\end{figure*}
\vspace{-1em}

\section{Discussion}
\vspace{-1em}
\subsection{Mechanism for Tension Onset}
\vspace{-1em}
\label{sec: Mechanism}
We have seen in the experimental results of Section~\ref{sec: Pressure_Field_Exp} that the observed onset of cavitation shortly after turnaround is accompanied by a sudden drop in pressure in the liquid to values below zero. We have also seen in Section~\ref{sec: Pressure_Field_Mod} that, by modeling the liquid column as a spring-mass system, we can capture the wave dynamics that may lead to the onset of tension in the liquid column. We now attempt to provide a complete explanation for the mechanism by which wave reflections may lead to tension in the liquid column and how it may explain the location of the bubble clusters.

Figure~\ref{fig: Pressure_Reflections} shows schematic representations of the mechanism of wave generation and reflection in our experiments. As illustrated in Fig.~\ref{fig: Pressure_Reflections}a, the abrupt slowing down of the free surface and subsequent compression of the gas volume sends a train of compression waves downward through the liquid. Before this compressive pulse, the piston's upward velocity ($v_\mathrm{p}$) is almost exactly equal to the water velocity ($v_\mathrm{w}$) and the pressure in the liquid column is relatively low (on the order of 0.1~MPa, as seen experimentally). As the train of waves reaches the piston, it imparts a downward momentum on the piston, thus slowing it down.

The large acoustic impedance difference between the water and the aluminum piston implies that acoustic waves will be reflected off of the piston surface. Due to the finite mass of the piston, the nature of that reflection will depend on the direction of the piston's momentum. Initially, the piston is moving upward when it gets hit by the first compressive waves. The wave slows down the piston, but the velocity of the piston relative to the water is still positive upward ($v_\mathrm{p}-\Delta v_\mathrm{p} > v_\mathrm{w}-\Delta v_\mathrm{w}$). Therefore, the initial compressive waves get reflected as compressive waves, as illustrated in Fig.~\ref{fig: Pressure_Reflections}b. Eventually, the piston reaches zero velocity, and the compressive waves start accelerating it downward. In this case, illustrated in Fig.~\ref{fig: Pressure_Reflections}c, the velocity of the piston relative to the water will be inverted, and therefore the piston will reflect compression waves as rarefaction waves, leading to a sudden decompression in the liquid.

The piston can therefore act as either a solid wall, which reflects compression waves as compression waves, or as a free surface, which reflects compression waves as rarefaction waves, depending on its momentum. The transition between these two types of reflection will depend on the magnitude of the pressure pulse, the mass and velocity of the piston, as well as the compressibility of the liquid medium. 

In addition to the reflection of the wave off of the piston, the wave will also reflect off of the free surface between the liquid column and the gas volume. Here, the type of reflection will depend on the magnitude of the wave as well as the magnitude of pressure and vertical size of the gas volume. Indeed, if the pressure in the gas volume is high and its volume small (as is the case near turnaround), a wave arriving at the free surface will \textit{ring-up} in the small air gap and reflect off of the solid wall at the top of the experiment (as seen in Fig.~\ref{fig: Pressure_Reflections}a and Fig.~\ref{fig: Model_delP}). If the pressure in the gas volume is low, the free surface will reflect compression waves as tension waves and vice-versa due to the large difference in acoustic impedance between the water and the gas.

As explained above, the magnitude and timescale of rarefaction waves in the liquid column will depend heavily on the velocity and mass of the piston as well as the strength of the initial compressive pulse. For tension to occur in the liquid column, a sufficiently strong rarefaction wave must be generated. This state of tension may be reached immediately after the rarefaction generated by the piston or may occur only after the reflection of this rarefaction wave leads to a further decrease in the liquid pressure. 
\vspace{-1em}

\subsection{Cavitation Regimes and Prediction}
We now compare experimental and modeling results over a wide range of parameters, with a focus on differentiating regimes of cavitation and of no cavitation, as well as testing the ability of the model to accurately predict cavitation. The experimental data used represents 129 experiments with varying peak pressures and velocities, obtained by changing the three independent parameters. The modeling data follows a similar range of parameters, however, due to the ease and speed of modeling, many more experiments are simulated that allow us to create better-refined phase maps.

\subsubsection{Cavitation Number}

\begin{figure}[b!]
\includegraphics[width = 0.5\textwidth]{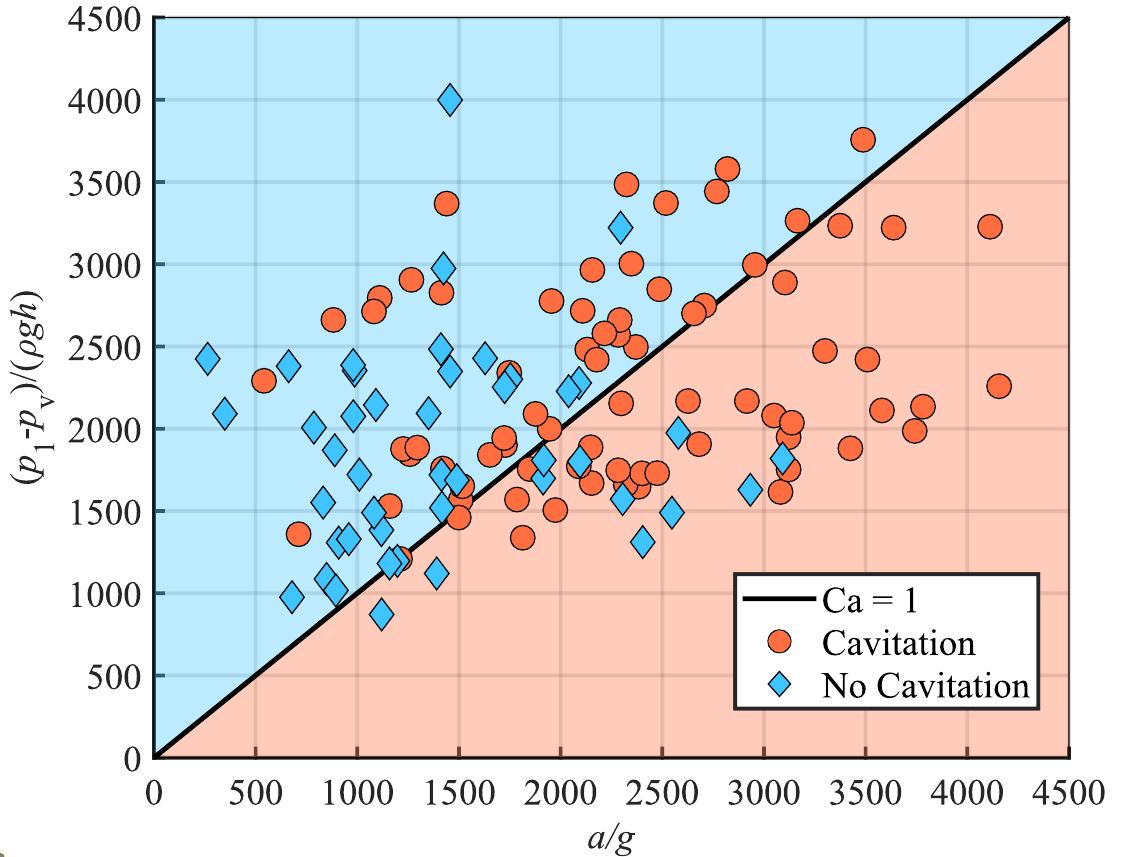}% Here is how to import EPS art
\caption{\label{fig: Cavinum_Phase_Map} Experimental phase map data showing the onset of cavitation on a plot where the $x$ and $y$ axes are the denominator and numerator of Eq.~\ref{eq: Ca2}, respectively. The line of $\mathrm{Ca}=1$ represents the theoretical boundary between cavitation (below the line) and no cavitation (above the line). These results show that this cavitation number is not able to predict cavitation onset for our experimental facility.}
\end{figure}
As mentioned in Section \ref{sec:Introduction}, previous work by other authors attempted to link cavitation with acceleration using rigid column theory\cite{Fatjo_2016, Pan_2017}. This was done through the cavitation number given in Eq.~\ref{eq: Ca2}, linking acceleration, column height, fluid density, and a reference static pressure to the onset of cavitation. This approach assumes that the timescale of acceleration ($\Delta t$) is large relative to the timescale of acoustic waves traveling the length of the liquid column ($t_\mathrm{wave}$), and therefore treats the liquid as incompressible.

\begin{figure*}[t!]
        \includegraphics[width = \textwidth]{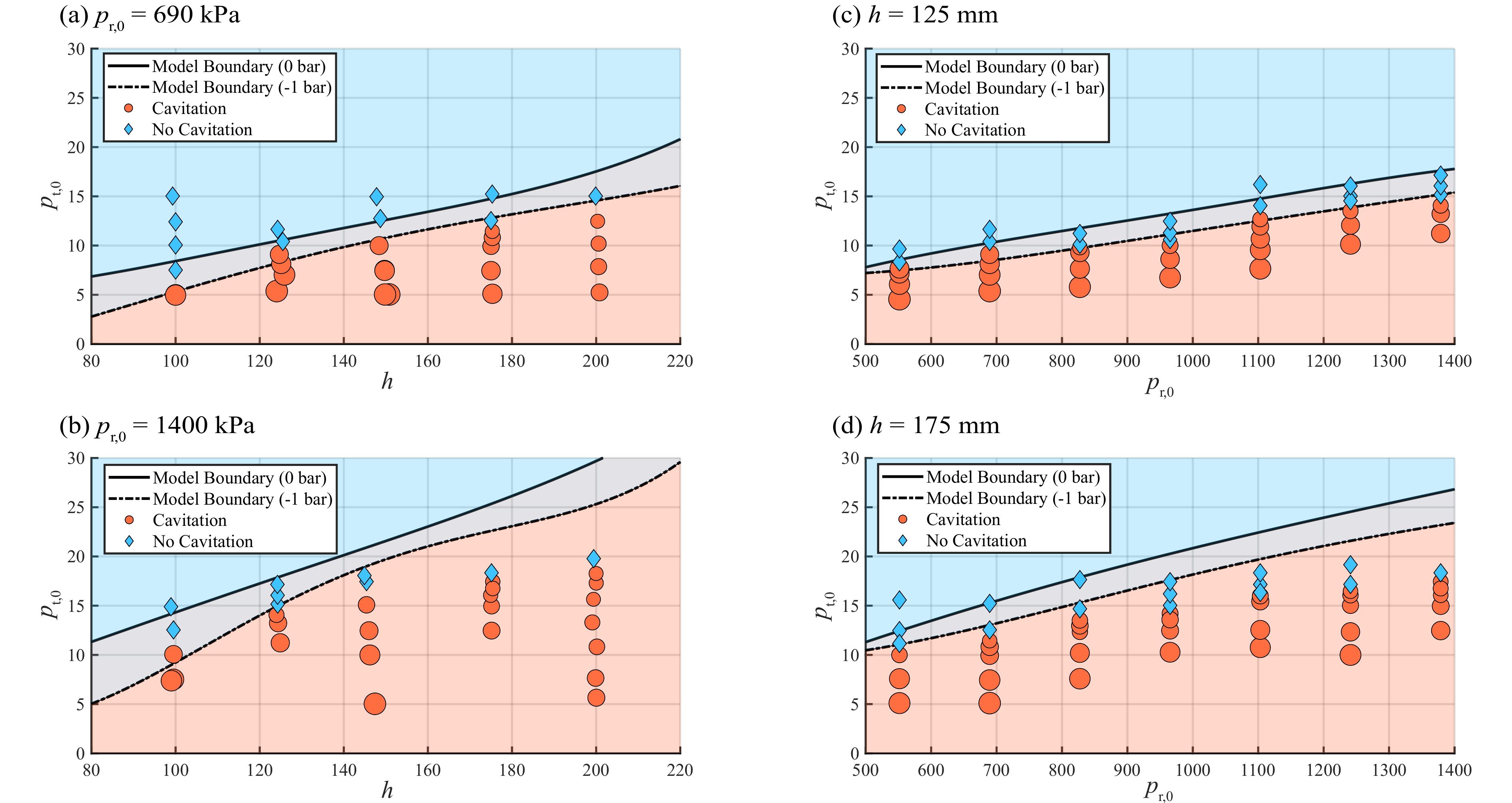}
\caption{\label{fig: Exp_Model_Phase_Maps} Phase maps showing modeling and experimental (markers) cavitation results. Plots a) and b) show the variation $h$ and $p_\mathrm{t,0}$ as the driving pressure $p_\mathrm{r,0}$ is maintained constant at 690~kPa and 1380~kPa, respectively. Plots c) and d) show the variation of $p_\mathrm{r,0}$ and $p_\mathrm{t,0}$ as $h$ is maintained constant at 125~mm and 175~mm, respectively. The size of red markers indicates the experimental size of the bubble cluster relative to the column height ($L_\mathrm{max}/h$).}
\end{figure*}
As seen in Section~\ref{sec: Mechanism}, our experiment is subjected to more complex dynamics and wave reflection mechanisms that are key to whether tension (and, therefore, cavitation) may occur in the liquid. In particular, the motion of the liquid column is such that $\Delta t \approx t_\mathrm{wave}$. In this condition, not only the magnitude of acceleration but also its timescale has a significant effect on the strength of the compression pulse. Therefore, the assumptions of rigid column theory (incompressibility and large $\Delta t$) are not valid in our apparatus.

Figure~\ref{fig: Cavinum_Phase_Map} shows a phase map of our experimental results where the $x$-axis is the denominator of Eq.~\ref{eq: Ca2} and the $y$-axis is its numerator. The acceleration is chosen as the maximum absolute acceleration of the liquid column, which occurs at the moment of turnaround. To be consistent with the literature and with our choice of acceleration, we define our reference static pressure $p_1$ as the pressure in the gas volume at the moment of turnaround. This reference pressure is physically meaningful as it accounts for the fact that the liquid is compressed at the moment of turnaround. The line of $\mathrm{Ca}=1$ is the theoretical limit below which cavitation is expected to occur. However, we see in Fig.~\ref{fig: Cavinum_Phase_Map} that this approach cannot predict the onset of cavitation for our experiments. This motivates the need for a new modeling tool that may correctly predict cavitation in a liquid column for more complex dynamics.

\subsubsection{Cavitation Prediction}
As mentioned in Section \ref{sec:Introduction}, tension in a liquid is a necessary but not sufficient condition for cavitation to occur. The modeling tool presented in Section~\ref{sec: Model_Dvlpmt} can predict tension in the liquid, but to use the model as a predictive tool for cavitation, a value of tension must be defined where we will consider pressures below a given threshold to represent cavitation prediction by the model. The determination of the exact negative pressure threshold below which the liquid column will cavitate can yield largely different results\cite{Williams_2004} and is beyond the scope of this paper. Our experimental results indicate that the cavitation we observe is likely heterogeneous nucleation, and therefore the magnitude of the negative pressure needed for cavitation is expected to be relatively low. To show that the model threshold can be adapted to any required value of negative pressure, we will show results for two thresholds: 0~bar and $-1$~bar.

\begin{figure*}[t!]
\includegraphics[width = \textwidth]{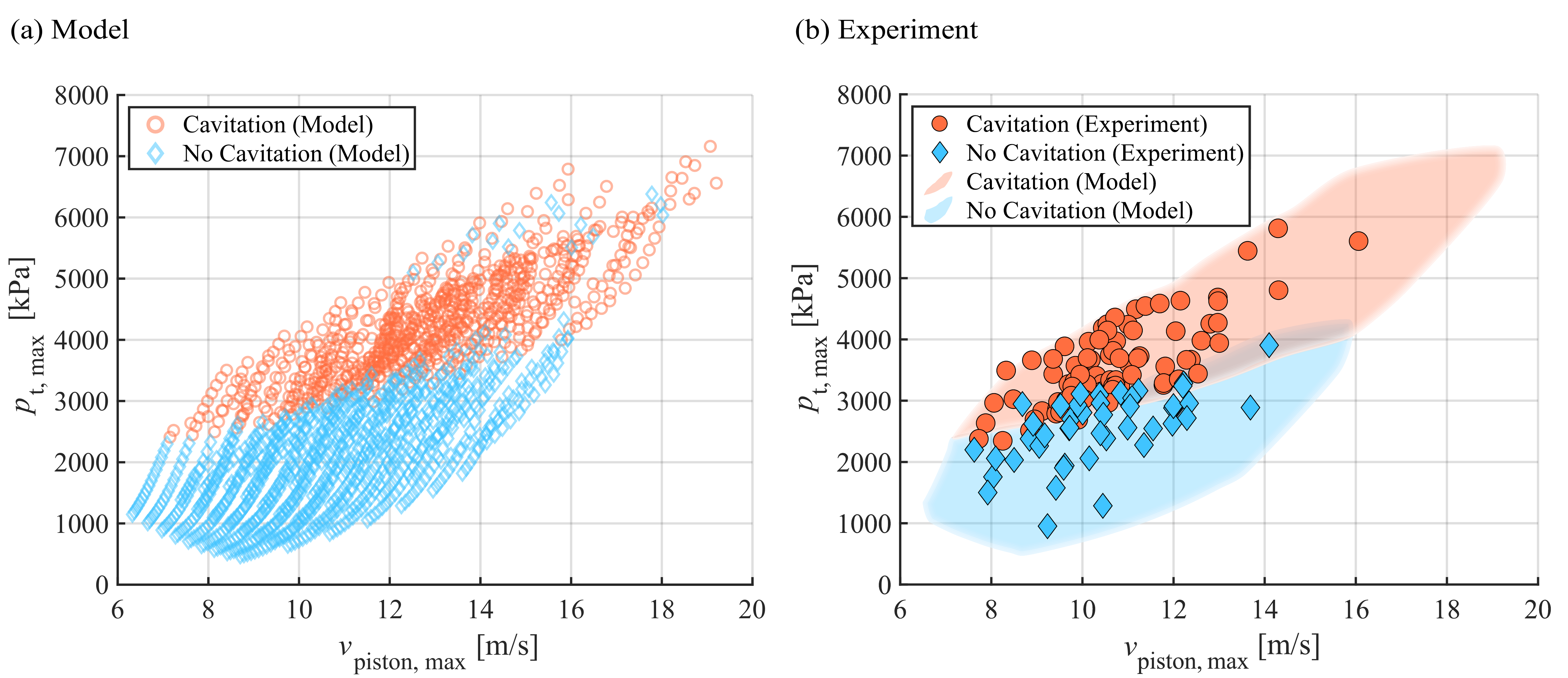}% Here is how to import EPS art
\caption{\label{fig: Exp_Model_Phase_Map_Phys} (a) Cavitation onset predicted by the model. (b) Cavitation onset seen experimentally. Both (a) and (b) show all experiments and simulations where each experiment is quantified by its maximum piston velocity and maximum test section pressure. The shaded regions in (b) represent the regions of the phase space where the model predicts mostly cavitation (in red) or no cavitation (in blue).}
\end{figure*}

Figure~\ref{fig: Exp_Model_Phase_Maps} reports the experimental and modeling results obtained for a range of experimental parameters. Blue diamond markers indicate experiments where cavitation was not observed, and red circles indicate experiments where cavitation was observed, with the size of the circle increasing with increasing bubble cluster size relative to the test section height ($L_\mathrm{max}/h$). The shaded blue and red regions correspond to regions in which the model predicts no cavitation and cavitation, respectively. The model's predicted cavitation boundary for the two thresholds of 0~bar and $-$1~bar (solid and dashed black lines, respectively) are also shown.

We can see from these results that the threshold of cavitation is highly sensitive to the initial test section pressure $p_\mathrm{t,0}$---cavitation occurs only for experiments where this is lowered significantly below atmospheric pressure. This is related to the observation made in Section \ref{sec: Pressure_Field_Exp} that a lower initial pressure in the gas volume will result in a more abrupt turnaround, leading to a larger pressure pulse and therefore greater likelihood of cavitation. We also find that the threshold of cavitation is sensitive to both an increase in test section height and driving pressure. An increase in driving pressure leads to higher peak velocities and accelerations, thus leading to a stronger compression and pressure pulse and a higher likelihood of cavitation. An increase in liquid column height will also increase the magnitude of the pressure pulse due to the added momentum of the column when it abruptly stops and turns around.

We find that the model boundaries of cavitation for both pressure thresholds (0~bar and $-1$~bar) fall close to what is seen experimentally. However, the model tends to slightly over-predict cavitation onset, particularly for experimental conditions of higher $p_\mathrm{r,0}$ and $h$ (Figs.~\ref{fig: Exp_Model_Phase_Maps}b and \ref{fig: Exp_Model_Phase_Maps}d). These experiments operate in a regime where higher accelerations are reached, leading to a more abrupt turnaround. For these conditions of high accelerations and velocities, energy losses in the system affect the dynamics of the piston. These losses, likely due to friction and heat transfer effects, are not captured by the model. This therefore explains the discrepancy between modeling and experimental cavitation that is observed in Fig.~\ref{fig: Exp_Model_Phase_Maps}. A more complete modeling approach for the piston motion would more accurately predict the boundary defining the onset of cavitation.

Although the phase maps shown in Fig.~\ref{fig: Exp_Model_Phase_Maps} are useful for our experimental setup, the parameters of $p_\mathrm{t,0}$ and $p_\mathrm{r,0}$ do not have physical significance outside the context of this experiment. The physical behavior can better be quantified using parameters relating to the velocities and pressures measured for each experiment. We found that the peak velocity ($v_\mathrm{piston, max}$) and peak pressure ($p_\mathrm{t,max}$) reached in a given experiment is a good way to parametrize the results. Figure~\ref{fig: Exp_Model_Phase_Map_Phys}a shows a phase map of all the simulated experiments based on these two parameters, and Fig.~\ref{fig: Exp_Model_Phase_Map_Phys}b shows experimental results on the same axes. Red circular markers indicate cavitation, blue diamond markers indicate no cavitation and full markers show experimental results whereas open markers show modeling results. 

Both sets of experimental and modeling results occupy a similar region of the phase space. The fact that many simulations were run permits us to have wider and better-resolved areas of the phase space. The area of the phase space where modeling results mostly show cavitation has been shaded in red and the region where it sees no cavitation is shaded in blue. We see that peak pressure and velocity define a boundary between experiments that cavitate and experiments that do not, and that the model and experiments are in good agreement. Some outliers can be seen and there is a slight overlap at the boundary between the two regions, but overall this confirms that the modeling tool can accurately predict the onset of cavitation in our experimental setup. 

Figure~\ref{fig: Exp_Model_Phase_Map_Phys} allows us to see that experiments where a higher peak pressure is reached (corresponding to a more abrupt turnaround) are more likely to cavitate. This is in agreement with the observations of Sec.~\ref{sec: Mechanism} which identified a link between the strength of the pressure pulse generated at turnaround and the onset and strength of cavitation. We also notice that an increase in peak velocity tends to make cavitation less likely for a given peak pressure. This is because higher peak velocities are reached for lower liquid heights: experiments with small $h$ have more travel distance and thus reach greater upward velocities. The lower mass of the water in these cases leads to a weaker pressure pulse generated at turnaround, and therefore a lower likelihood of cavitation. These effects are accurately modeled by our modeling tool, thus making it a useful tool for cavitation prediction in impulsively accelerated liquid columns.
\vspace{-1em}

\section{Conclusion}
We have presented a new experimental apparatus for cavitation in an impulsively accelerated liquid column and developed a modeling tool that allows us to simulate the experiment and predict the onset of cavitation. The experiments presented used water columns between 100--200~mm accelerated at up to 90\,000~m/s$^2$, reaching pressures in the liquid of up to 6~MPa. 

The experimental measurements using high-speed imagery and piezoelectric pressure transducers revealed that cavitation onset occurs within about 1~ms after the column comes to a stop and reaches peak pressure and acceleration. The compression and ensuing pressure drop in the column can lead to negative pressures in the liquid that are responsible for the onset of cavitation. Pressure measurements and images revealed that cavitation may occur away from the top and bottom surfaces of the liquid due to the reflection of compression and rarefaction waves in the liquid column. The cavitation was observed to occur in the form of a dense bubble cluster, in which growing bubbles tend to merge and collapse as a single bubble. It was also observed that the collapse of the bubble cluster generated a strong pressure increase emanating from the point of collapse, which can reflect off of free surfaces and induce secondary cavitation onset in the liquid.

The modeling tool developed was used to reproduce experiments in a way that replicated both the motion and the pressure field within the liquid column. This pressure field and its gradient illustrated the ability of the model to accurately predict the wave dynamics and to reproduce the mechanism by which tension occurs in the liquid. The model was also run for a large range of parameters to identify a boundary for regimes of cavitation onset. The predicted boundary was compared to the experimental results, and it was found that the model could accurately predict the onset of cavitation and could therefore reliably be used as a prediction tool for cavitation. 

The modeling results showed that predicting the pressure field in a liquid column can be done accurately with a spring-mass approach. This could in principle be generalized to any dynamical system in which a liquid column might be expected to cavitate. The spring-mass model can readily be expressed for a system with different dynamics (e.g., drop-weight or tube-arrest experiment) and should be able to capture the wave dynamics and the onset of tension. The accuracy of the cavitation prediction is dependent on a reliable model for the motion that the liquid column is subjected to. In the present study, the modeling tool used to predict the motion of the liquid column was found to be less accurate at higher accelerations. Future work will focus on modeling the dynamics of the system more accurately to refine the cavitation prediction boundary.

Since this study is the first to provide a simple modeling tool to relate the dynamics of an impulsively accelerated liquid column to its pressure field and tension in the column, its focus has mainly been on accurately predicting the onset of tension and potential cavitation. However, the model currently is not able to model the growth and collapse of bubble clusters in the liquid and the ensuing pressure field. To resolve this shortcoming, the model could be coupled with a Rayleigh-Plesset-type model that could accurately predict the growth and collapse of cavitation bubbles in the liquid column given the pressure field using an approach similar to those used by Denner and Schenke\cite{Denner_2023} and Bryngelson et al\cite{BRYNGELSON2019137}. 

\vspace{-1.5em}
\begin{acknowledgments}
\vspace{-1em}
The authors thank Andreas Hoffman, Ramnarine Harihar, Meisam Aghajani, Mathieu Beauchesne, and Sam Minter for assistance with the machining and design of the experimental apparatus. The authors also acknowledge helpful discussions with Nick Sirmas, Piotr Forysinski, and Fabian Denner. This work was supported by the Canadian Natural Science and Engineering Research Council (grant nr. ALLRP/553986-2020), General Fusion, and the McGill Summer Undergraduate Research in Engineering program.
\end{acknowledgments}
\vspace{-1.0em}

\section*{Author Declaration}

\subsection*{Conflict of Interest Statement} 
\vspace{-1.0em}

The authors have no conflicts to disclose.
\vspace{-1.0em}
\subsection*{Author Contributions}
\vspace{-1.0em}
\textbf{Taj Sobral:} Conceptualization (equal); Data Curation (lead); Formal Analysis (equal); Investigation (equal); Methodology (equal); Project Administration (equal); Software (lead); Validation (lead); Visualization (lead); Writing -- original draft (lead); Writing -- Review \& Editing (lead). \textbf{John Kokkalis:} Conceptualization (equal); Data Curation (support); Formal Analysis (equal); Investigation (equal); Methodology (equal); Software (support); Visualization (support).   \textbf{Kay Romann:} Methodology (support); Validation (support). \textbf{Jovan Nedić:} Conceptualization (equal); Project Administration (equal); Supervision (equal); Writing -- original draft (support); Writing -- Review \& Editing (support).
\textbf{Andrew J. Higgins:} Conceptualization (equal); Project Administration (equal); Supervision (equal); Writing -- original draft (support); Writing -- Review \& Editing (support).

\vspace{-1.0em}
\section*{Data Availability Statement}

The data that support the findings of this study are available from the corresponding author upon reasonable request.
\begin{figure}[b]
\includegraphics[width = 0.5\textwidth]{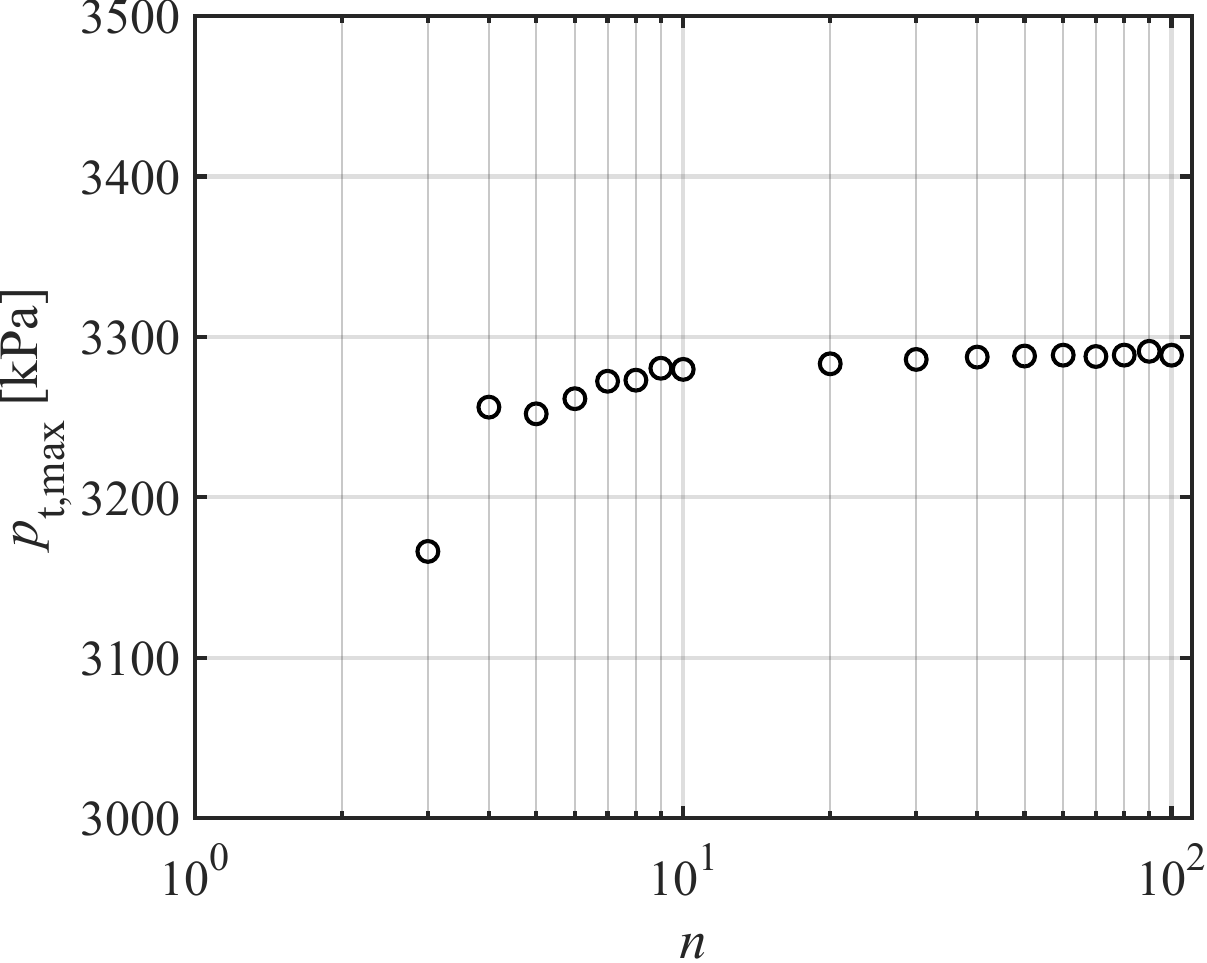}% Here is how to import EPS art
\caption{\label{fig: Model_Convergence_Pt} Convergence plot of modeling results showing the value of $p_\mathrm{t, max}$ for mass numbers from $n=3$ to $100$.}
\end{figure}
\appendix
\vspace{-1.0em}
\section{Model Convergence}

As mentioned above, the model treats the water as a system of discrete point masses connected by springs. We can increase the spatial resolution by increasing the number of masses in the liquid column with every additional mass adding two equations to our ODE system and therefore increases computing time. Figure~\ref{fig: Model_Convergence_Pt} shows a convergence plot of the maximum pressure in the test section ($p_\mathrm{t, max}$) for a given experiment as the model is refined from $n=3$ to $100$, with convergence occurring for $n=10$. We use $n=10$ to run the large phase maps shown in the paper because the low computing time allows us to simulate more experiments and produce better-resolved phase maps. To provide better spatial resolution, the results for individual modeled experiments shown in the paper use $n=50$. 
\vspace{-1.0em}

\section{Additional Experiments}
\label{App: Exp}
\begin{figure}[h!]
    \centering
    \includegraphics[width=0.85\linewidth]{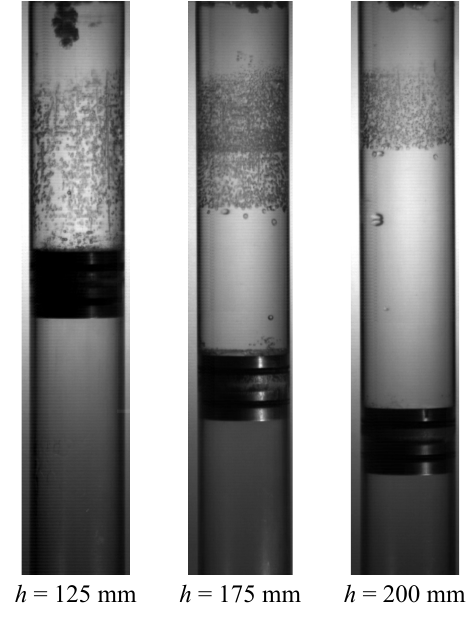}
    \caption{Images of cavitation onset for typical experiments at different heights (multimedia available online). The experimental conditions are: (left) $h = 125$ mm, $p_\mathrm{t,max} = 4.0$~MPa, and $v_\mathrm{p,max} = 10.1$~m/s, (middle) $h = 175$ mm, $p_\mathrm{t,max} = 3.7$~MPa, and $v_\mathrm{p,max} = 8.89$~m/s, (right) $h = 200$ mm, $p_\mathrm{t,max} = 4.0$~MPa, and $v_\mathrm{p,max} = 10.4$~m/s.}
    \label{fig: App_B}
\end{figure}

In addition to the experimental data shown in the main body of the paper, we provide here some additional experimental data in order to illustrate the effect of water height on cavitation onset. Figure~\ref{fig: App_B} shows images of typical experiments in which cavitation occurs at different heights of the water column (multimedia available online). We can observe that the height of the cavitation zone relative to the total height of the column is not constant across experiments. However, the images (and associated videos) show that the behavior of cavitation onset and collapse is qualitatively equivalent for all column heights.
\nocite{*}
%\bibliography{Manuscript}% Produces the bibliography via BibTeX.

\begin{thebibliography}{30}%
\makeatletter
\providecommand \@ifxundefined [1]{%
 \@ifx{#1\undefined}
}%
\providecommand \@ifnum [1]{%
 \ifnum #1\expandafter \@firstoftwo
 \else \expandafter \@secondoftwo
 \fi
}%
\providecommand \@ifx [1]{%
 \ifx #1\expandafter \@firstoftwo
 \else \expandafter \@secondoftwo
 \fi
}%
\providecommand \natexlab [1]{#1}%
\providecommand \enquote  [1]{``#1''}%
\providecommand \bibnamefont  [1]{#1}%
\providecommand \bibfnamefont [1]{#1}%
\providecommand \citenamefont [1]{#1}%
\providecommand \href@noop [0]{\@secondoftwo}%
\providecommand \href [0]{\begingroup \@sanitize@url \@href}%
\providecommand \@href[1]{\@@startlink{#1}\@@href}%
\providecommand \@@href[1]{\endgroup#1\@@endlink}%
\providecommand \@sanitize@url [0]{\catcode `\\12\catcode `\$12\catcode `\&12\catcode `\#12\catcode `\^12\catcode `\_12\catcode `\%12\relax}%
\providecommand \@@startlink[1]{}%
\providecommand \@@endlink[0]{}%
\providecommand \url  [0]{\begingroup\@sanitize@url \@url }%
\providecommand \@url [1]{\endgroup\@href {#1}{\urlprefix }}%
\providecommand \urlprefix  [0]{URL }%
\providecommand \Eprint [0]{\href }%
\providecommand \doibase [0]{http://dx.doi.org/}%
\providecommand \selectlanguage [0]{\@gobble}%
\providecommand \bibinfo  [0]{\@secondoftwo}%
\providecommand \bibfield  [0]{\@secondoftwo}%
\providecommand \translation [1]{[#1]}%
\providecommand \BibitemOpen [0]{}%
\providecommand \bibitemStop [0]{}%
\providecommand \bibitemNoStop [0]{.\EOS\space}%
\providecommand \EOS [0]{\spacefactor3000\relax}%
\providecommand \BibitemShut  [1]{\csname bibitem#1\endcsname}%
\let\auto@bib@innerbib\@empty
%</preamble>
\bibitem [{\citenamefont {Brennen}(2013)}]{Brennen_2013}%
  \BibitemOpen
  \bibfield  {author} {\bibinfo {author} {\bibfnamefont {C.~E.}\ \bibnamefont {Brennen}},\ }\href@noop {} {\emph {\bibinfo {title} {Cavitation and Bubble Dynamics}}}\ (\bibinfo  {publisher} {Cambridge University Press},\ \bibinfo {year} {2013})\BibitemShut {NoStop}%
\bibitem [{\citenamefont {Brennen}(2011)}]{Brennen_2011}%
  \BibitemOpen
  \bibfield  {author} {\bibinfo {author} {\bibfnamefont {C.~E.}\ \bibnamefont {Brennen}},\ }\href@noop {} {\emph {\bibinfo {title} {Hydrodynamics of Pumps}}}\ (\bibinfo  {publisher} {Cambridge University Press},\ \bibinfo {year} {2011})\BibitemShut {NoStop}%
\bibitem [{\citenamefont {Cole}(1965)}]{Cole_1965}%
  \BibitemOpen
  \bibfield  {author} {\bibinfo {author} {\bibfnamefont {R.~H.}\ \bibnamefont {Cole}},\ }\href@noop {} {\emph {\bibinfo {title} {Underwater Explosions}}}\ (\bibinfo  {publisher} {Dover},\ \bibinfo {year} {1965})\BibitemShut {NoStop}%
\bibitem [{\citenamefont {Roovers}\ \emph {et~al.}(2019)\citenamefont {Roovers}, \citenamefont {Segers}, \citenamefont {Lajoinie}, \citenamefont {Deprez}, \citenamefont {Versluis}, \citenamefont {De~Smedt},\ and\ \citenamefont {Lentacker}}]{Roovers_2019}%
  \BibitemOpen
  \bibfield  {author} {\bibinfo {author} {\bibfnamefont {S.}~\bibnamefont {Roovers}}, \bibinfo {author} {\bibfnamefont {T.}~\bibnamefont {Segers}}, \bibinfo {author} {\bibfnamefont {G.}~\bibnamefont {Lajoinie}}, \bibinfo {author} {\bibfnamefont {J.}~\bibnamefont {Deprez}}, \bibinfo {author} {\bibfnamefont {M.}~\bibnamefont {Versluis}}, \bibinfo {author} {\bibfnamefont {S.~C.}\ \bibnamefont {De~Smedt}}, \ and\ \bibinfo {author} {\bibfnamefont {I.}~\bibnamefont {Lentacker}},\ }\bibfield  {title} {\enquote {\bibinfo {title} {The role of ultrasound-driven microbubble dynamics in drug delivery: From microbubble fundamentals to clinical translation},}\ }\href {\doibase 10.1021/acs.langmuir.8b03779} {\bibfield  {journal} {\bibinfo  {journal} {Langmuir}\ }\textbf {\bibinfo {volume} {35}},\ \bibinfo {pages} {10173--10191} (\bibinfo {year} {2019})}\BibitemShut {NoStop}%
\bibitem [{\citenamefont {Veilleux}\ and\ \citenamefont {Shepherd}(2018)}]{Veilleux_2018}%
  \BibitemOpen
  \bibfield  {author} {\bibinfo {author} {\bibfnamefont {J.-C.}\ \bibnamefont {Veilleux}}\ and\ \bibinfo {author} {\bibfnamefont {E.~J.}\ \bibnamefont {Shepherd}},\ }\bibfield  {title} {\enquote {\bibinfo {title} {Pressure and stress transients in autoinjector devices},}\ }\href {\doibase 10.1007/s13346-018-0568-7} {\bibfield  {journal} {\bibinfo  {journal} {Drug Delivery and Translational Research}\ }\textbf {\bibinfo {volume} {8}},\ \bibinfo {pages} {1238--1253} (\bibinfo {year} {2018})}\BibitemShut {NoStop}%
\bibitem [{\citenamefont {Trevena}(1984)}]{Trevena_1984}%
  \BibitemOpen
  \bibfield  {author} {\bibinfo {author} {\bibfnamefont {D.~H.}\ \bibnamefont {Trevena}},\ }\bibfield  {title} {\enquote {\bibinfo {title} {Cavitation and the generation of tension in liquids},}\ }\href {\doibase 10.1088/0022-3727/17/11/003} {\bibfield  {journal} {\bibinfo  {journal} {Journal of Physics D: Applied Physics}\ }\textbf {\bibinfo {volume} {17}},\ \bibinfo {pages} {2139} (\bibinfo {year} {1984})}\BibitemShut {NoStop}%
\bibitem [{\citenamefont {Knapp}(1958)}]{Knapp_1958}%
  \BibitemOpen
  \bibfield  {author} {\bibinfo {author} {\bibfnamefont {R.~T.}\ \bibnamefont {Knapp}},\ }\bibfield  {title} {\enquote {\bibinfo {title} {{Cavitation and Nuclei}},}\ }\href {\doibase 10.1115/1.4012694} {\bibfield  {journal} {\bibinfo  {journal} {Transactions of the American Society of Mechanical Engineers}\ }\textbf {\bibinfo {volume} {80}},\ \bibinfo {pages} {1315--1324} (\bibinfo {year} {1958})}\BibitemShut {NoStop}%
\bibitem [{\citenamefont {Pan}\ \emph {et~al.}(2017)\citenamefont {Pan}, \citenamefont {Kiyama}, \citenamefont {Tagawa}, \citenamefont {Daily}, \citenamefont {Thomson}, \citenamefont {Hurd},\ and\ \citenamefont {Truscott}}]{Pan_2017}%
  \BibitemOpen
  \bibfield  {author} {\bibinfo {author} {\bibfnamefont {Z.}~\bibnamefont {Pan}}, \bibinfo {author} {\bibfnamefont {A.}~\bibnamefont {Kiyama}}, \bibinfo {author} {\bibfnamefont {Y.}~\bibnamefont {Tagawa}}, \bibinfo {author} {\bibfnamefont {D.~J.}\ \bibnamefont {Daily}}, \bibinfo {author} {\bibfnamefont {S.~L.}\ \bibnamefont {Thomson}}, \bibinfo {author} {\bibfnamefont {R.}~\bibnamefont {Hurd}}, \ and\ \bibinfo {author} {\bibfnamefont {T.~T.}\ \bibnamefont {Truscott}},\ }\bibfield  {title} {\enquote {\bibinfo {title} {Cavitation onset caused by acceleration},}\ }\href {\doibase 10.1073/pnas.1702502114} {\bibfield  {journal} {\bibinfo  {journal} {Proceedings of the National Academy of Sciences}\ }\textbf {\bibinfo {volume} {114}},\ \bibinfo {pages} {8470--8474} (\bibinfo {year} {2017})}\BibitemShut {NoStop}%
\bibitem [{\citenamefont {Garcia-Atance~Fatjo}(2016)}]{Fatjo_2016}%
  \BibitemOpen
  \bibfield  {author} {\bibinfo {author} {\bibfnamefont {G.}~\bibnamefont {Garcia-Atance~Fatjo}},\ }\bibfield  {title} {\enquote {\bibinfo {title} {New dimensionless number to predict cavitation in accelerated fluid},}\ }\href {\doibase 10.2495/CMEM-V4-N4-484-492} {\bibfield  {journal} {\bibinfo  {journal} {International Journal of Computational Methods and Experimental Measurements}\ }\textbf {\bibinfo {volume} {4}},\ \bibinfo {pages} {484--492} (\bibinfo {year} {2016})}\BibitemShut {NoStop}%
\bibitem [{\citenamefont {Berthelot}(1850)}]{Berthelot_1849}%
  \BibitemOpen
  \bibfield  {author} {\bibinfo {author} {\bibfnamefont {M.}~\bibnamefont {Berthelot}},\ }\bibfield  {title} {\enquote {\bibinfo {title} {Sur quelques ph{\'e}nom{\`e}nes de dilatation forc{\'e}e des liquides},}\ }\href {https://books.google.ca/books?id=JERS0AEACAAJ} {\bibfield  {journal} {\bibinfo  {journal} {Annales de Physique et de Chimie}\ }\textbf {\bibinfo {volume} {30}},\ \bibinfo {pages} {232--237} (\bibinfo {year} {1850})}\BibitemShut {NoStop}%
\bibitem [{\citenamefont {Trevena}(1978)}]{Trevena_1978}%
  \BibitemOpen
  \bibfield  {author} {\bibinfo {author} {\bibfnamefont {D.~H.}\ \bibnamefont {Trevena}},\ }\bibfield  {title} {\enquote {\bibinfo {title} {Marcelin {B}erthelot's first publication in 1850, on the subjection of liquids to tension},}\ }\href {\doibase 10.1080/00033797800200131} {\bibfield  {journal} {\bibinfo  {journal} {Annals of Science}\ }\textbf {\bibinfo {volume} {35}},\ \bibinfo {pages} {45--54} (\bibinfo {year} {1978})}\BibitemShut {NoStop}%
\bibitem [{\citenamefont {Frenkel}(1955)}]{Frenkel_1955}%
  \BibitemOpen
  \bibfield  {author} {\bibinfo {author} {\bibfnamefont {J.}~\bibnamefont {Frenkel}},\ }\href@noop {} {\emph {\bibinfo {title} {Kinetic {T}heory of {L}iquids}}}\ (\bibinfo  {publisher} {Dover Publications, Inc.},\ \bibinfo {year} {1955})\BibitemShut {NoStop}%
\bibitem [{\citenamefont {Bull}(1956)}]{Bull_1956}%
  \BibitemOpen
  \bibfield  {author} {\bibinfo {author} {\bibfnamefont {T.~H.}\ \bibnamefont {Bull}},\ }\bibfield  {title} {\enquote {\bibinfo {title} {The tensile strengths of liquids under dynamic loading},}\ }\href {\doibase 10.1080/14786435608238088} {\bibfield  {journal} {\bibinfo  {journal} {The Philosophical Magazine: A Journal of Theoretical Experimental and Applied Physics}\ }\textbf {\bibinfo {volume} {1}},\ \bibinfo {pages} {153--165} (\bibinfo {year} {1956})}\BibitemShut {NoStop}%
\bibitem [{\citenamefont {Skripov}(1974)}]{Skripov}%
  \BibitemOpen
  \bibfield  {author} {\bibinfo {author} {\bibfnamefont {V.~P.}\ \bibnamefont {Skripov}},\ }\href@noop {} {\emph {\bibinfo {title} {Metastable {L}iquids}}}\ (\bibinfo  {publisher} {J. Wiley},\ \bibinfo {year} {1974})\BibitemShut {NoStop}%
\bibitem [{\citenamefont {Couzens}\ and\ \citenamefont {Trevena}(1974)}]{Couzens_1974}%
  \BibitemOpen
  \bibfield  {author} {\bibinfo {author} {\bibfnamefont {D.~C.~F.}\ \bibnamefont {Couzens}}\ and\ \bibinfo {author} {\bibfnamefont {D.~H.}\ \bibnamefont {Trevena}},\ }\bibfield  {title} {\enquote {\bibinfo {title} {Tensile failure of liquids under dynamic stressing},}\ }\href {\doibase 10.1088/0022-3727/7/16/315} {\bibfield  {journal} {\bibinfo  {journal} {Journal of Physics D: Applied Physics}\ }\textbf {\bibinfo {volume} {7}},\ \bibinfo {pages} {2277} (\bibinfo {year} {1974})}\BibitemShut {NoStop}%
\bibitem [{\citenamefont {Richards}, \citenamefont {Trevena},\ and\ \citenamefont {Edwards}(1980)}]{Richards_1980}%
  \BibitemOpen
  \bibfield  {author} {\bibinfo {author} {\bibfnamefont {B.~E.}\ \bibnamefont {Richards}}, \bibinfo {author} {\bibfnamefont {D.~H.}\ \bibnamefont {Trevena}}, \ and\ \bibinfo {author} {\bibfnamefont {D.~H.}\ \bibnamefont {Edwards}},\ }\bibfield  {title} {\enquote {\bibinfo {title} {Cavitation experiments using a water shock tube},}\ }\href {\doibase 10.1088/0022-3727/13/7/027} {\bibfield  {journal} {\bibinfo  {journal} {Journal of Physics D: Applied Physics}\ }\textbf {\bibinfo {volume} {13}},\ \bibinfo {pages} {1315} (\bibinfo {year} {1980})}\BibitemShut {NoStop}%
\bibitem [{\citenamefont {Joseph}(1998)}]{JOSEPH_1998}%
  \BibitemOpen
  \bibfield  {author} {\bibinfo {author} {\bibfnamefont {D.~D.}\ \bibnamefont {Joseph}},\ }\bibfield  {title} {\enquote {\bibinfo {title} {Cavitation and the state of stress in a flowing liquid},}\ }\href {\doibase 10.1017/S0022112098001530} {\bibfield  {journal} {\bibinfo  {journal} {Journal of Fluid Mechanics}\ }\textbf {\bibinfo {volume} {366}},\ \bibinfo {pages} {367–378} (\bibinfo {year} {1998})}\BibitemShut {NoStop}%
\bibitem [{\citenamefont {Williams}\ and\ \citenamefont {Williams}(2004)}]{Williams_2004}%
  \BibitemOpen
  \bibfield  {author} {\bibinfo {author} {\bibfnamefont {P.~R.}\ \bibnamefont {Williams}}\ and\ \bibinfo {author} {\bibfnamefont {R.~L.}\ \bibnamefont {Williams}},\ }\bibfield  {title} {\enquote {\bibinfo {title} {Cavitation and the tensile strength of liquids under dynamic stressing},}\ }\href {\doibase 10.1080/00268970412331292786} {\bibfield  {journal} {\bibinfo  {journal} {Molecular Physics}\ }\textbf {\bibinfo {volume} {102}},\ \bibinfo {pages} {2091--2102} (\bibinfo {year} {2004})}\BibitemShut {NoStop}%
\bibitem [{\citenamefont {Huneault}\ and\ \citenamefont {Higgins}(2019)}]{Huneault_2019}%
  \BibitemOpen
  \bibfield  {author} {\bibinfo {author} {\bibfnamefont {J.}~\bibnamefont {Huneault}}\ and\ \bibinfo {author} {\bibfnamefont {A.~J.}\ \bibnamefont {Higgins}},\ }\bibfield  {title} {\enquote {\bibinfo {title} {{Shock wave induced cavitation of silicone oils}},}\ }\href {\doibase 10.1063/1.5093028} {\bibfield  {journal} {\bibinfo  {journal} {Journal of Applied Physics}\ }\textbf {\bibinfo {volume} {125}},\ \bibinfo {pages} {245903} (\bibinfo {year} {2019})}\BibitemShut {NoStop}%
\bibitem [{\citenamefont {Xu}\ \emph {et~al.}(2020)\citenamefont {Xu}, \citenamefont {Liu}, \citenamefont {Zuo},\ and\ \citenamefont {Pan}}]{Xu_2020}%
  \BibitemOpen
  \bibfield  {author} {\bibinfo {author} {\bibfnamefont {P.}~\bibnamefont {Xu}}, \bibinfo {author} {\bibfnamefont {S.}~\bibnamefont {Liu}}, \bibinfo {author} {\bibfnamefont {Z.}~\bibnamefont {Zuo}}, \ and\ \bibinfo {author} {\bibfnamefont {Z.~H.}\ \bibnamefont {Pan}},\ }\bibfield  {title} {\enquote {\bibinfo {title} {On the criteria of large cavitation bubbles in a tube during a transient process},}\ }\href {https://api.semanticscholar.org/CorpusID:222290806} {\bibfield  {journal} {\bibinfo  {journal} {Journal of Fluid Mechanics}\ }\textbf {\bibinfo {volume} {913}} (\bibinfo {year} {2020})}\BibitemShut {NoStop}%
\bibitem [{\citenamefont {Qi-Dai}\ and\ \citenamefont {Long}(2004)}]{Chen_2004}%
  \BibitemOpen
  \bibfield  {author} {\bibinfo {author} {\bibfnamefont {C.}~\bibnamefont {Qi-Dai}}\ and\ \bibinfo {author} {\bibfnamefont {W.}~\bibnamefont {Long}},\ }\bibfield  {title} {\enquote {\bibinfo {title} {Production of large size single transient cavitation bubbles with tube arrest method},}\ }\href {\doibase 10.1088/1009-1963/13/4/028} {\bibfield  {journal} {\bibinfo  {journal} {Chinese Physics}\ }\textbf {\bibinfo {volume} {13}},\ \bibinfo {pages} {564} (\bibinfo {year} {2004})}\BibitemShut {NoStop}%
\bibitem [{\citenamefont {Laberge}(2019)}]{Laberge}%
  \BibitemOpen
  \bibfield  {author} {\bibinfo {author} {\bibfnamefont {M.}~\bibnamefont {Laberge}},\ }\bibfield  {title} {\enquote {\bibinfo {title} {Magnetized target fusion with a spherical tokamak},}\ }\href {https://doi.org/10.1007/s10894-018-0180-3} {\bibfield  {journal} {\bibinfo  {journal} {Journal of Fusion Energy}\ }\textbf {\bibinfo {volume} {35}},\ \bibinfo {pages} {199203} (\bibinfo {year} {2019})}\BibitemShut {NoStop}%
\bibitem [{\citenamefont {Sobral}\ \emph {et~al.}(2023)\citenamefont {Sobral}, \citenamefont {Kokkalis}, \citenamefont {Higgins}, \citenamefont {Nedić}, \citenamefont {Sirmas},\ and\ \citenamefont {Forysinski}}]{Sobral_2023}%
  \BibitemOpen
  \bibfield  {author} {\bibinfo {author} {\bibfnamefont {T.}~\bibnamefont {Sobral}}, \bibinfo {author} {\bibfnamefont {J.}~\bibnamefont {Kokkalis}}, \bibinfo {author} {\bibfnamefont {A.}~\bibnamefont {Higgins}}, \bibinfo {author} {\bibfnamefont {J.}~\bibnamefont {Nedić}}, \bibinfo {author} {\bibfnamefont {N.}~\bibnamefont {Sirmas}}, \ and\ \bibinfo {author} {\bibfnamefont {P.}~\bibnamefont {Forysinski}},\ }\bibfield  {title} {\enquote {\bibinfo {title} {Acoustic pulse generation in collapse of a liquid cavity},}\ }in\ \href@noop {} {\emph {\bibinfo {booktitle} {Proceedings of the 34th International Symposium on Shock Waves (ISSW34), Daegu, Korea, July 16-21}}}\ (\bibinfo {year} {2023})\BibitemShut {NoStop}%
\bibitem [{\citenamefont {Kokkalis}(2023)}]{Kokkalis_2023}%
  \BibitemOpen
  \bibfield  {author} {\bibinfo {author} {\bibfnamefont {J.}~\bibnamefont {Kokkalis}},\ }\emph {\bibinfo {title} {Onset of Cavitation in a Dynamically Loaded Piston-Cylinder}},\ \href {https://escholarship.mcgill.ca/concern/theses/q524jv097} {Master's thesis},\ \bibinfo  {school} {McGill University} (\bibinfo {year} {2023})\BibitemShut {NoStop}%
\bibitem [{\citenamefont {Strand}\ \emph {et~al.}(2006)\citenamefont {Strand}, \citenamefont {Goosman}, \citenamefont {Martinez}, \citenamefont {Whitworth},\ and\ \citenamefont {Kuhlow}}]{Strand_2006}%
  \BibitemOpen
  \bibfield  {author} {\bibinfo {author} {\bibfnamefont {O.~T.}\ \bibnamefont {Strand}}, \bibinfo {author} {\bibfnamefont {D.~R.}\ \bibnamefont {Goosman}}, \bibinfo {author} {\bibfnamefont {C.}~\bibnamefont {Martinez}}, \bibinfo {author} {\bibfnamefont {T.~L.}\ \bibnamefont {Whitworth}}, \ and\ \bibinfo {author} {\bibfnamefont {W.~W.}\ \bibnamefont {Kuhlow}},\ }\bibfield  {title} {\enquote {\bibinfo {title} {Compact system for high-speed velocimetry using heterodyne techniques},}\ }\href {\doibase 10.1063/1.2336749} {\bibfield  {journal} {\bibinfo  {journal} {Review of Scientific Instruments}\ }\textbf {\bibinfo {volume} {35}},\ \bibinfo {pages} {083108} (\bibinfo {year} {2006})}\BibitemShut {NoStop}%
\bibitem [{\citenamefont {Dolan}(2010)}]{Dolan}%
  \BibitemOpen
  \bibfield  {author} {\bibinfo {author} {\bibfnamefont {D.~H.}\ \bibnamefont {Dolan}},\ }\bibfield  {title} {\enquote {\bibinfo {title} {Accuracy and precision in photonic Doppler velocimetry},}\ }\href {\doibase 10.1063/1.3429257} {\bibfield  {journal} {\bibinfo  {journal} {Review of Scientific Instruments}\ }\textbf {\bibinfo {volume} {81}},\ \bibinfo {pages} {053905} (\bibinfo {year} {2010})},\ \Eprint {http://arxiv.org/abs/https://pubs.aip.org/aip/rsi/article-pdf/doi/10.1063/1.3429257/13678551/053905\_1\_online.pdf} {https://pubs.aip.org/aip/rsi/article-pdf/doi/10.1063/1.3429257/13678551/053905\_1\_online.pdf} \BibitemShut {NoStop}%
\bibitem [{\citenamefont {Richer}\ and\ \citenamefont {Hurmuzlu}(2000)}]{Richer_2000}%
  \BibitemOpen
  \bibfield  {author} {\bibinfo {author} {\bibfnamefont {E.}~\bibnamefont {Richer}}\ and\ \bibinfo {author} {\bibfnamefont {Y.}~\bibnamefont {Hurmuzlu}},\ }\bibfield  {title} {\enquote {\bibinfo {title} {A high performance pneumatic force actuator system: Part i—nonlinear mathematical model},}\ }\href {https://api.semanticscholar.org/CorpusID:14121748} {\bibfield  {journal} {\bibinfo  {journal} {Journal of Dynamic Systems Measurement and Control-transactions of The {ASME}}\ }\textbf {\bibinfo {volume} {122}},\ \bibinfo {pages} {416--425} (\bibinfo {year} {2000})}\BibitemShut {NoStop}%
\bibitem [{\citenamefont {Shepherd}\ and\ \citenamefont {Inaba}(2010)}]{Shepherd_2010}%
  \BibitemOpen
  \bibfield  {author} {\bibinfo {author} {\bibfnamefont {J.~E.}\ \bibnamefont {Shepherd}}\ and\ \bibinfo {author} {\bibfnamefont {K.}~\bibnamefont {Inaba}},\ }\enquote {\bibinfo {title} {Shock loading and failure of fluid-filled tubular structures},}\ in\ \href {\doibase 10.1007/978-1-4419-0446-1_6} {\emph {\bibinfo {booktitle} {Dynamic Failure of Materials and Structures}}},\ \bibinfo {editor} {edited by\ \bibinfo {editor} {\bibfnamefont {A.}~\bibnamefont {Shukla}}, \bibinfo {editor} {\bibfnamefont {G.}~\bibnamefont {Ravichandran}}, \ and\ \bibinfo {editor} {\bibfnamefont {Y.~D.}\ \bibnamefont {Rajapakse}}}\ (\bibinfo  {publisher} {Springer US},\ \bibinfo {address} {Boston, MA},\ \bibinfo {year} {2010})\ pp.\ \bibinfo {pages} {153--190}\BibitemShut {NoStop}%
\bibitem [{\citenamefont {Denner}\ and\ \citenamefont {Schenke}(2023)}]{Denner_2023}%
  \BibitemOpen
  \bibfield  {author} {\bibinfo {author} {\bibfnamefont {F.}~\bibnamefont {Denner}}\ and\ \bibinfo {author} {\bibfnamefont {S.}~\bibnamefont {Schenke}},\ }\bibfield  {title} {\enquote {\bibinfo {title} {Modeling acoustic emissions and shock formation of cavitation bubbles},}\ }\href {\doibase 10.1063/5.0131930} {\bibfield  {journal} {\bibinfo  {journal} {Physics of Fluids}\ }\textbf {\bibinfo {volume} {35}},\ \bibinfo {pages} {012114} (\bibinfo {year} {2023})}\BibitemShut {NoStop}%
\bibitem [{\citenamefont {Bryngelson}, \citenamefont {Schmidmayer},\ and\ \citenamefont {Colonius}(2019)}]{BRYNGELSON2019137}%
  \BibitemOpen
  \bibfield  {author} {\bibinfo {author} {\bibfnamefont {S.~H.}\ \bibnamefont {Bryngelson}}, \bibinfo {author} {\bibfnamefont {K.}~\bibnamefont {Schmidmayer}}, \ and\ \bibinfo {author} {\bibfnamefont {T.}~\bibnamefont {Colonius}},\ }\bibfield  {title} {\enquote {\bibinfo {title} {A quantitative comparison of phase-averaged models for bubbly, cavitating flows},}\ }\href {\doibase https://doi.org/10.1016/j.ijmultiphaseflow.2019.03.028} {\bibfield  {journal} {\bibinfo  {journal} {International Journal of Multiphase Flow}\ }\textbf {\bibinfo {volume} {115}},\ \bibinfo {pages} {137--143} (\bibinfo {year} {2019})}\BibitemShut {NoStop}%
\end{thebibliography}
\providecommand{\noopsort}[1]{}\providecommand{\singleletter}[1]{#1}%

\end{document}